\documentclass[journal=jacsat, manuscript=article, layout=onecolumn]{achemso}

\makeatletter
\def\@fnsymbol#1{\ensuremath{\ifcase#1\or \ddagger\or *\or \dagger\or	\mathsection\or \mathparagraph\or **\or \dagger\dagger\or \ddagger\ddagger \else\@ctrerr\fi}}
\makeatother

\usepackage{graphicx}
\usepackage{color,amssymb,amsmath}
\usepackage[normalem]{ulem}
\usepackage{hyperref}
\usepackage{lipsum}
\usepackage{comment}

\mciteErrorOnUnknownfalse

\newcommand{\onlinecite}[1]{\hspace{-1 ex} \nocite{#1}~\citenum{#1}} 

\newcommand{\Isw}{I_{\mathrm{SW}}}
\newcommand{\PhiR}{\mathit{\Phi}_{\mathrm{R}}}
\newcommand{\PhiL}{\mathit{\Phi}_{\mathrm{L}}}
\newcommand{\Phio}{\mathit{\Phi}_{\mathrm{0}}}
\newcommand{\IL}{I_{\mathrm{L}}}
\newcommand{\IR}{I_{\mathrm{R}}}
\newcommand{\Ial}{I_{\mathrm{Al}}}

\title{Demonstration of nonlocal Josephson effect in Andreev molecules}

\author{D. Z. Haxell}
\altaffiliation{These authors contributed equally.}
\affiliation{IBM Research Europe---Zurich, S\"aumerstrasse 4, 8803 R\"uschlikon, Switzerland}

\author{M. Coraiola}
\altaffiliation{These authors contributed equally.}
\affiliation{IBM Research Europe---Zurich, S\"aumerstrasse 4, 8803 R\"uschlikon, Switzerland}

\author{M. Hinderling}
\affiliation{IBM Research Europe---Zurich, S\"aumerstrasse 4, 8803 R\"uschlikon, Switzerland}

\author{S. C. ten Kate}
\affiliation{IBM Research Europe---Zurich, S\"aumerstrasse 4, 8803 R\"uschlikon, Switzerland}

\author{D. Sabonis}
\affiliation{IBM Research Europe---Zurich, S\"aumerstrasse 4, 8803 R\"uschlikon, Switzerland}

\author{A.\ E.\ Svetogorov}
\affiliation{Fachbereich Physik, Universit\"at Konstanz, D-78457 Konstanz, Germany}

\author{W.\ Belzig}
\affiliation{Fachbereich Physik, Universit\"at Konstanz, D-78457 Konstanz, Germany}

\author{E. Cheah}
\affiliation{Solid State Physics Laboratory, ETH Z\"urich, 8093 Z\"urich, Switzerland}

\author{F. Krizek}
\alsoaffiliation[Present address: ]{Institute of Physics, Czech Academy of Sciences, 162 00 Prague, Czech Republic}
\affiliation{IBM Research Europe---Zurich, S\"aumerstrasse 4, 8803 R\"uschlikon, Switzerland}
\affiliation{Solid State Physics Laboratory, ETH Z\"urich, 8093 Z\"urich, Switzerland}

\author{R. Schott}
\affiliation{Solid State Physics Laboratory, ETH Z\"urich, 8093 Z\"urich, Switzerland}

\author{W. Wegscheider}
\affiliation{Solid State Physics Laboratory, ETH Z\"urich, 8093 Z\"urich, Switzerland}

\author{F. Nichele}
\email{fni@zurich.ibm.com}
\affiliation{IBM Research Europe---Zurich, S\"aumerstrasse 4, 8803 R\"uschlikon, Switzerland}

\date{\today}

\begin{document}

\begin{abstract}
We perform switching current measurements of planar Josephson junctions (JJs) coupled by a common superconducting electrode, with independent control over the two superconducting phase differences. We observe an anomalous phase shift in the current--phase relation of a JJ as a function of gate voltage or phase difference in the second JJ. This demonstrates a nonlocal Josephson effect, and the implementation of a $\varphi_0$-junction which is tunable both electrostatically and magnetically. The anomalous phase shift was larger for shorter distances between the JJs and vanished for distances much longer than the superconducting coherence length. Results are consistent with the hybridization of ABSs, leading to the formation of an Andreev molecule. Our devices constitute a realization of a tunable superconducting phase source, and could enable new coupling schemes for hybrid quantum devices.
\end{abstract}


\maketitle
The development of high-quality hybrid superconductor--semiconductor materials over the last decade enabled new possibilities in superconducting electronics and quantum computing~\cite{Krogstrup2015,Chang2015,Shabani2016}. In particular, Andreev bound states (ABSs)~\cite{Andreev1964,Beenakker1991,Furusaki1991} arising in superconductor--semiconductor--superconductor Josephson junctions (JJs)~\cite{Pillet2010,Bretheau2013b,Tosi2019,Nichele2020} offer functionalities not attainable in metallic JJs. A prominent example is the electrostatic tuning of the critical current~\cite{Doh2005,Xiang2006,Shabani2016}, which allows for JJ field-effect transistors~\cite{Gheewala1980, Clark1980, Kleinsasser1989, Wen2019}, voltage-tunable superconducting qubits~\cite{Larsen2015,deLange2015,Casparis2018,PitaVidal2020}, resonators~\cite{Casparis2019,Sardashti2020} and amplifiers~\cite{Butseraen2022,Sarkar2022,Phan2023}. Moreover, the interplay between ABSs, spin--orbit interaction and Zeeman fields results in non-reciprocal switching currents~\cite{Baumgartner2022,Turini2022,Gupta2023,Matsuo2023c} and anomalous phase offsets, or $\varphi_0$-junctions~\cite{Buzdin2003,Buzdin2008,Yokoyama2014,Bergeret2015,Szombati2016,Hart2017,Assouline2019,Mayer2020,Strambini2020}, with applications in superconducting electronics and spintronics \cite{Linder2015}.

A yet largely unexplored possibility offered by superconductor--semiconductor hybrids is the engineering of Andreev molecules from the hybridization of spatially overlapping ABSs~\cite{Pillet2019,Kornich2019,Pillet2020,Kornich2020}. Predicted to arise in JJs coupling over length scales comparable to the superconducting coherence length, Andreev molecules offer a promising platform to realize $\varphi_0$-junctions \cite{Pillet2019} and novel manipulation and coupling schemes for Andreev qubits~\cite{Kornich2019}. Experimental studies of ABS hybridization focused on two-terminal quantum dots~\cite{Su2017,Kurtossy2021} and quantum dot chains~\cite{Dvir2023}. Recently, engineering of Andreev molecules was demonstrated in open, multiply connected geometries~\cite{Coraiola2023} and laterally coupled JJs~\cite{Matsuo2022,Matsuo2023}. Measurements of the switching current in double InAs nanowires revealed a nonlocal Josephson effect~\cite{Matsuo2022}, however the device geometry did not allow measurements of phase shifts in the current--phase relation (CPR). 

\begin{figure*}
	\includegraphics[width=\textwidth]{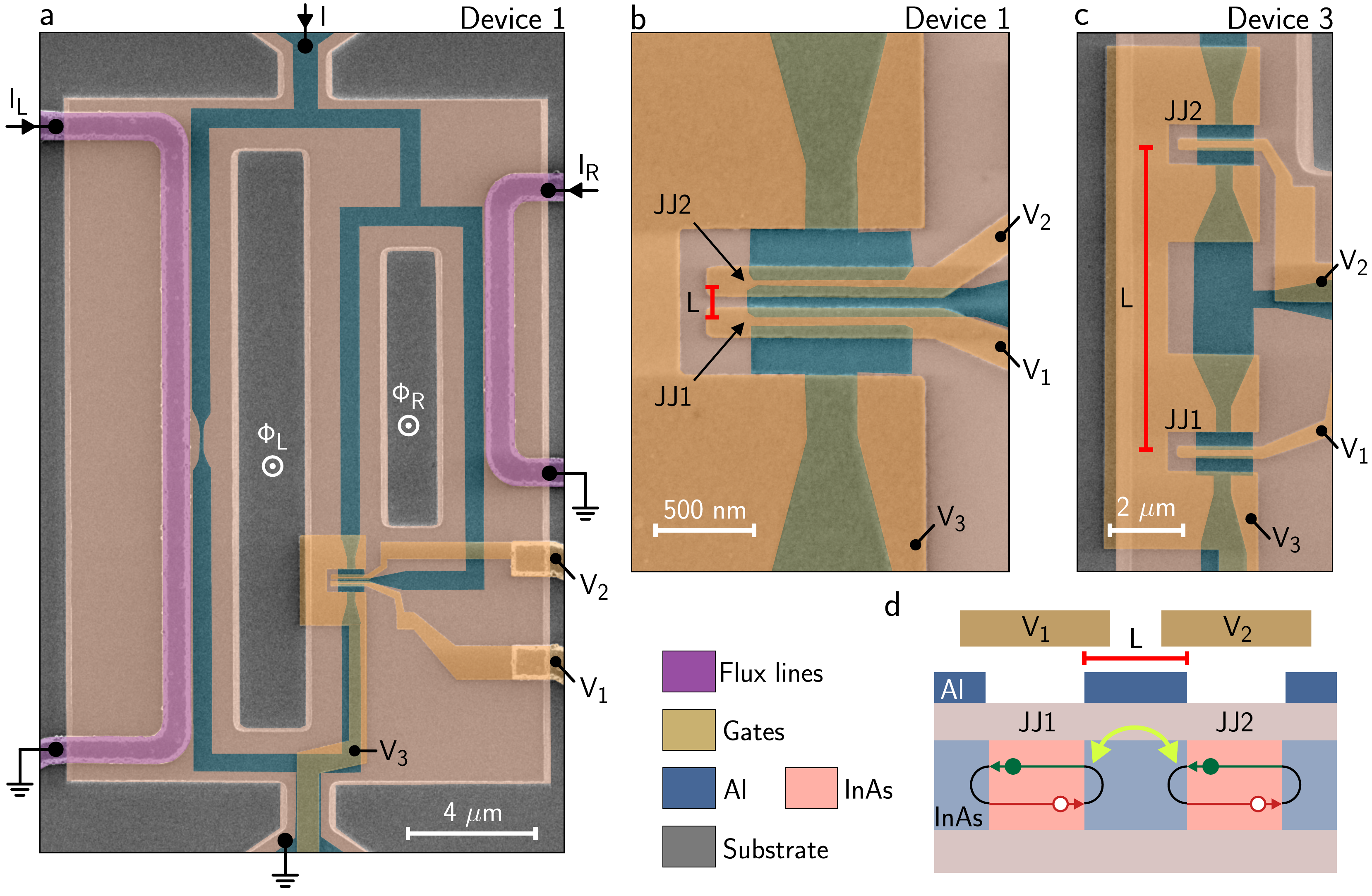}
	\caption{Devices under study and measurement setup. (a) False-colored scanning electron micrograph of a sample identical to Device~1, together with a measurement schematics. The substrate is shown in gray, the exposed III-V semiconductor in pink, the epitaxial Al in blue, gates in yellow and flux lines in purple. (b) Zoom-in of (a) around the Josephson junctions (JJs). The distance between the junctions is $L=150~\mathrm{nm}$. (c) Similar to (b), but for Device~3, which has $L=4~\mathrm{\mu m}$. (d) Schematic cross-section (not to scale) of the JJs sharing a common electrode of length $L$. Andreev bound states originating from the two JJs, spatially extended over distances in excess of $L$, overlap, and hybridize forming an Andreev molecule.}
	\label{fig1}
\end{figure*}

Here we demonstrate generation and electrical tuning of an anomalous phase shift in planar JJs that share a mesoscopic superconducting electrode. Inspired by Ref.~\onlinecite{Pillet2019}, we realized devices consisting of two JJs sharing a common electrode and embedded in a superconducting double-loop geometry. The expected phase anomaly was ascribed~\cite{Pillet2020} to the interplay between two distinct Cooper pair transfer mechanisms at $\varphi_1 = \varphi_2$ and at $\varphi_1 = -\varphi_2$, namely double crossed Andreev reflection (dCAR) and double elastic cotunneling (dEC), respectively. Differently from the proposal of Ref.~\onlinecite{Pillet2020}, which considered semiconductor nanowires, we used planar JJs containing several ABSs. Such devices are lithographically defined with a top-down approach and, tuning their geometry, allow for large switching currents.

The double-loop geometry allowed us to independently tune the superconducting phase differences $\varphi_1$ and $\varphi_2$ across the two JJs, named JJ1 and JJ2, respectively. Furthermore, it allowed us to characterize the CPR of JJ1 for different values of $\varphi_2$. A coupling between JJ1 and JJ2 manifested as distorted and phase-shifted CPR of JJ1 for $\varphi_2\neq(0,\pi)$. Such a coupled system realizes the nonlocal Josephson effect: an anomalous phase shift (or, equivalently, an anomalous supercurrent at zero phase difference) was nonlocally induced in JJ1 by the phase difference across JJ2. Varying the length $L$ of the common superconducting electrode, we observed that the phase shift was larger for small $L$, and vanished for $L$ much longer than the superconducting coherence length. Our observations are consistent with the hybridization of ABSs originating from the two JJs, resulting in the formation of an Andreev molecule. As the superconducting phase offset is electrically tuned by a current flowing in a flux line or a voltage applied to a gate, our devices constitute a demonstration of tunable superconducting phase source.  Our findings open up new avenues for the design and implementation of advanced nanoscale quantum devices with enhanced controllability and functionality.

Experiments were performed on four devices (Devices~1 to 4) defined in the same epitaxial heterostructure of InAs and Al~\cite{Shabani2016,Cheah2023}, measured in a dilution refrigerator with a base temperature below $10~\mathrm{mK}$. Figure~\ref{fig1}(a) shows a false-colored scanning electron micrograph of Device~1, indicating the exposed InAs (pink), the epitaxial Al (blue), the gate electrodes (yellow) and the flux-bias lines (purple). Devices consisted of a small superconducting loop, threaded by the flux $\PhiR$, embedded in the arm of a large superconducting loop, threaded by the flux $\PhiL$. The region where the two loops merged [bottom right in Fig.~\ref{fig1}(a)] is shown in Fig.~\ref{fig1}(b). Here, three Al leads defined two nominally identical JJs which shared a central Al electrode of length $L$. We label the bottom and top junction in Fig.~\ref{fig1}(b) JJ1 and JJ2, respectively. In each JJ, the width of the Al electrodes was $800~\mathrm{nm}$ and the length of the junction was $40~\mathrm{nm}$. From the junctions geometry, we estimate between $40$ to $100$ transverse modes to be present, for typical values of electron sheet density (see Supporting Information for details). Both JJ1 and JJ2 were controlled by gate electrodes, energized by voltages $V_1$ and $V_2$, respectively. A third gate, at voltage $V_3$, was set to $-3~\mathrm{V}$ throughout the experiment to prevent parallel conducting paths. A narrow Al constriction was defined on the left arm of the large Al loop [see Fig.~\ref{fig1}(a)]. This constriction limited the maximum supercurrent flowing in the device, making switching current measurements feasible without warming up the apparatus. Devices~1 to 3 differed exclusively by the parameter $L$, which was $150$, $400$ and $4000~\mathrm{nm}$, respectively. Device~4 was lithographically identical to Device~1 and is presented in the Supporting Information, together with further details on the heterostructure and sample fabrication. 

The measurement setup used to measure switching currents is schematically depicted in Fig.~\ref{fig1}(a). After compensating for a global magnetic field offset using a vector magnet, local magnetic fluxes were generated by applying slowly varying currents $\IL$ and $\IR$ in the flux lines. Switching currents of the entire device were obtained by ramping the current $I$ from zero to around $35~\mathrm{\mu A}$ (depending on the device) with a repetition rate of $133~\mathrm{Hz}$, and detecting when the four-terminal voltage drop $V$ exceeded a threshold. As planar devices are characterized by an intrinsically large spread of their switching current~\cite{Haxell2023}, we averaged the results over $16$ ramps for each data point. In all of our devices, JJ1 had a maximum supercurrent of approximately $450~\mathrm{nA}$, while the Al constrictions consistently showed switching currents of $\Ial\sim34~\mathrm{\mu A}$, independent of $\PhiL$ and $\PhiR$. Due to the large asymmetry between the arms of the device, the switching current $\Isw$ of the right arm was obtained by subtracting $\Ial$ from the switching current of the entire device. Further details on the measurement setup are discussed in the Supporting Information.

Our devices allowed an independent tuning of the phase differences across the two junctions. The phase difference $\varphi_1$ across JJ1 was tuned by the magnetic flux impinging within the perimeter of the device as {$\varphi_1=2\pi(\PhiL+\PhiR)/\Phio$}, where $\Phio$ is the superconducting magnetic flux quantum. The phase difference $\varphi_2$ across JJ2 was instead tuned by the magnetic flux $\PhiR$. However, as JJ2 was bypassed by a large stripe of Al, $\varphi_2$ was not expected to affect $\Isw$, unless a nonlocal Josephson effect was present; in the absence of coupling between JJ1 and JJ2, $\Isw$ would simply represent the CPR of JJ1.

We further note that the finite geometric and kinetic inductances of the superconducting loops were too small to significantly distort the CPR of JJ1 and JJ2. Similarly, coupling between the two loops mediated by a shared inductance was negligible. The inductances of the inner and outer loops were $105~\mathrm{pH}$ and $200~\mathrm{pH}$, respectively, while their common Al segment [right side of Fig.~\ref{fig1}(a)] had an inductance of $58~\mathrm{pH}$. Such values are significantly smaller than the Josephson inductances of JJ1 and JJ2, which were always greater than $800~\mathrm{pH}$. The absence of inductive coupling was experimentally confirmed by results obtained on Device~3, as discussed below.  

\begin{figure*}
	\includegraphics[width=\textwidth]{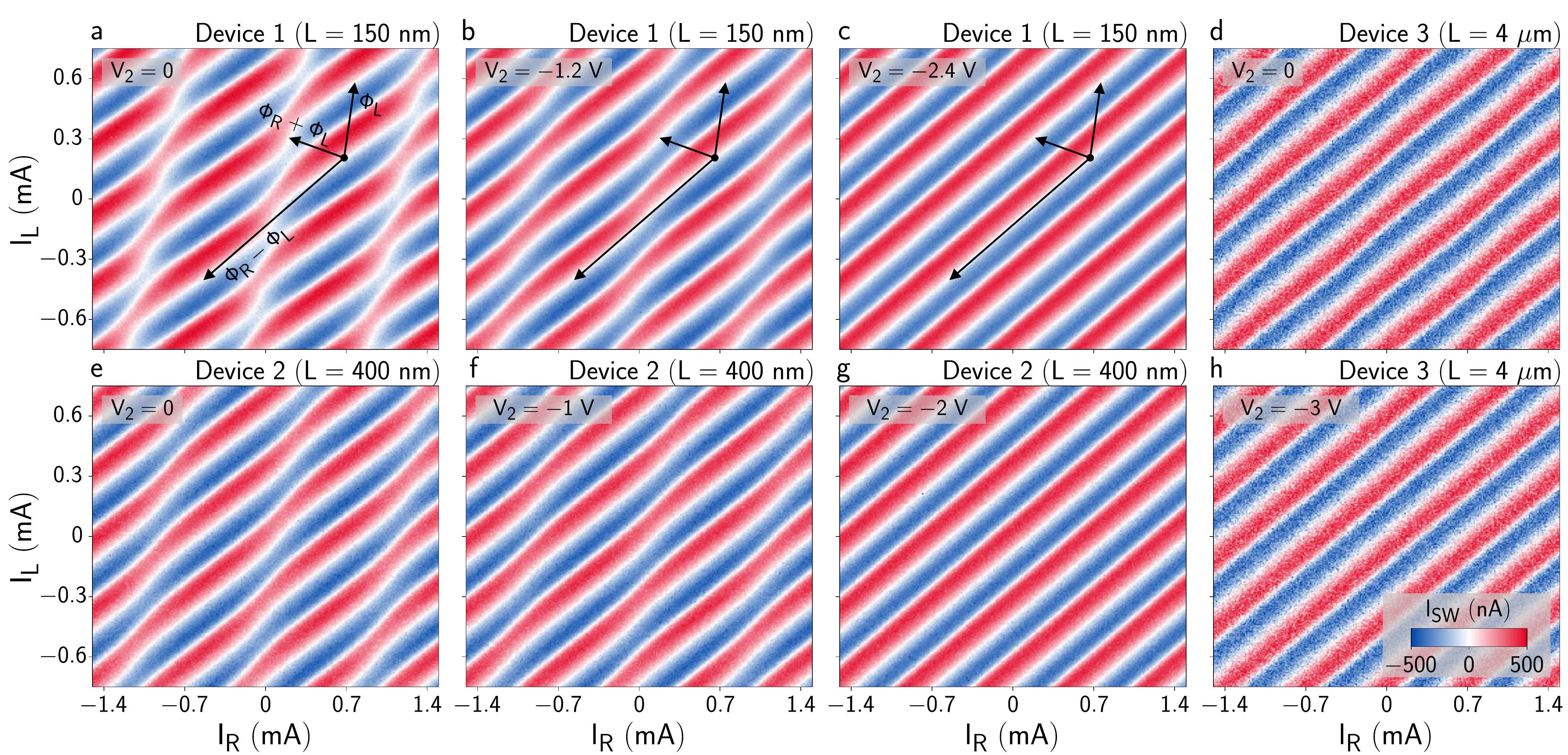}
	\caption{Phase-dependent supercurrent and evidence of nonlocal Josephson effect. (a-c) Switching current $\Isw$ for Device 1 after subtraction of the switching current of the Al constriction $\Ial$, for $V_1=0$ and $V_2=0$, $-1.2$ and $-2.4~\mathrm{V}$, respectively. Black arrows indicate the direction $\PhiL+\PhiR$, which is the direction of maximal modulation of $\varphi_1$, $\PhiR-\PhiL$, which is the direction over which $\varphi_1$ is constant, and $\PhiL$, which is the direction over which $\varphi_2$ is constant. (d) $\Isw$ in Device~3 for $V_1=V_2=0$. (e-g) Same as (a-c), but for Device~2. (h) $\Isw$ in Device~3 for $V_1=V_2=-3~\mathrm{V}$.}
	\label{fig2}
\end{figure*}

Figure~\ref{fig2}(a) shows the switching current $\Isw$ of Device~1 as a function of $\IL$ and $\IR$ and performed with $V_1=0$ and $V_2=0$ (both JJs open). Clear supercurrent oscillations were present, which depended on both $\IL$ and $\IR$. The vector $\PhiL+\PhiR$ in Fig.~\ref{fig2}(a) shows the direction over which $\varphi_1$ is expected to be maximally modulated. The vectors $\PhiR-\PhiL$ and $\PhiL$ indicate the directions over which $\varphi_1$ and $\varphi_2$ are constant, respectively. Details on the definition of such vectors are presented in the Supporting Information. Clear modulations of the supercurrent were observed along $\PhiR-\PhiL$, confirming a coupling between JJ1 and JJ2. Figures~\ref{fig2}(b, c) show equivalent measurements performed after setting $V_2=-1.2~\mathrm{V}$ and $V_2=-2.4~\mathrm{V}$, respectively. As $V_2$ was set more negative, JJ2 was depleted and the supercurrent oscillations became more and more regular, until supercurrent modulations were completely suppressed along $\PhiR-\PhiL$ [Fig.~\ref{fig2}(c)] and JJ1 displayed a conventional forward-skewed CPR. Further, we note that $V_2$ did not alter the maximum switching current amplitude, confirming the absence of trivial electrostatic coupling between the gate of JJ2 and JJ1. Supercurrent measurements for Device~2, which had $L=400~\mathrm{nm}$, are shown in Figs.~\ref{fig2} (e-g) for $V_1=0$ and varying $V_2$. Anomalous phase modulations were still present in the supercurrent oscillations of Fig.~\ref{fig2}(e), despite being significantly weaker than in Fig.~\ref{fig2}(a). Setting $V_2=-2~\mathrm{V}$ [Fig.~\ref{fig2}(g)] suppressed any remaining phase modulation, resulting in a conventional CPR as in Fig.~\ref{fig2}(c). Finally, Figs.~\ref{fig2}(d,~h) show measurements performed in Device~3, with $L=4~\mathrm{\mu m}$. In this case, phase modulations never occurred along $\PhiR-\PhiL$, neither for $V_2=0$ [Fig.~\ref{fig2}(d)] nor when $V_2=-3~\mathrm{V}$ [Fig.~\ref{fig2}(h)], demonstrating the absence of coupling in well-separated JJs and that contributions of loop inductance and circulating currents to phase shifts were negligible.

\begin{figure*}
	\includegraphics[width=\textwidth]{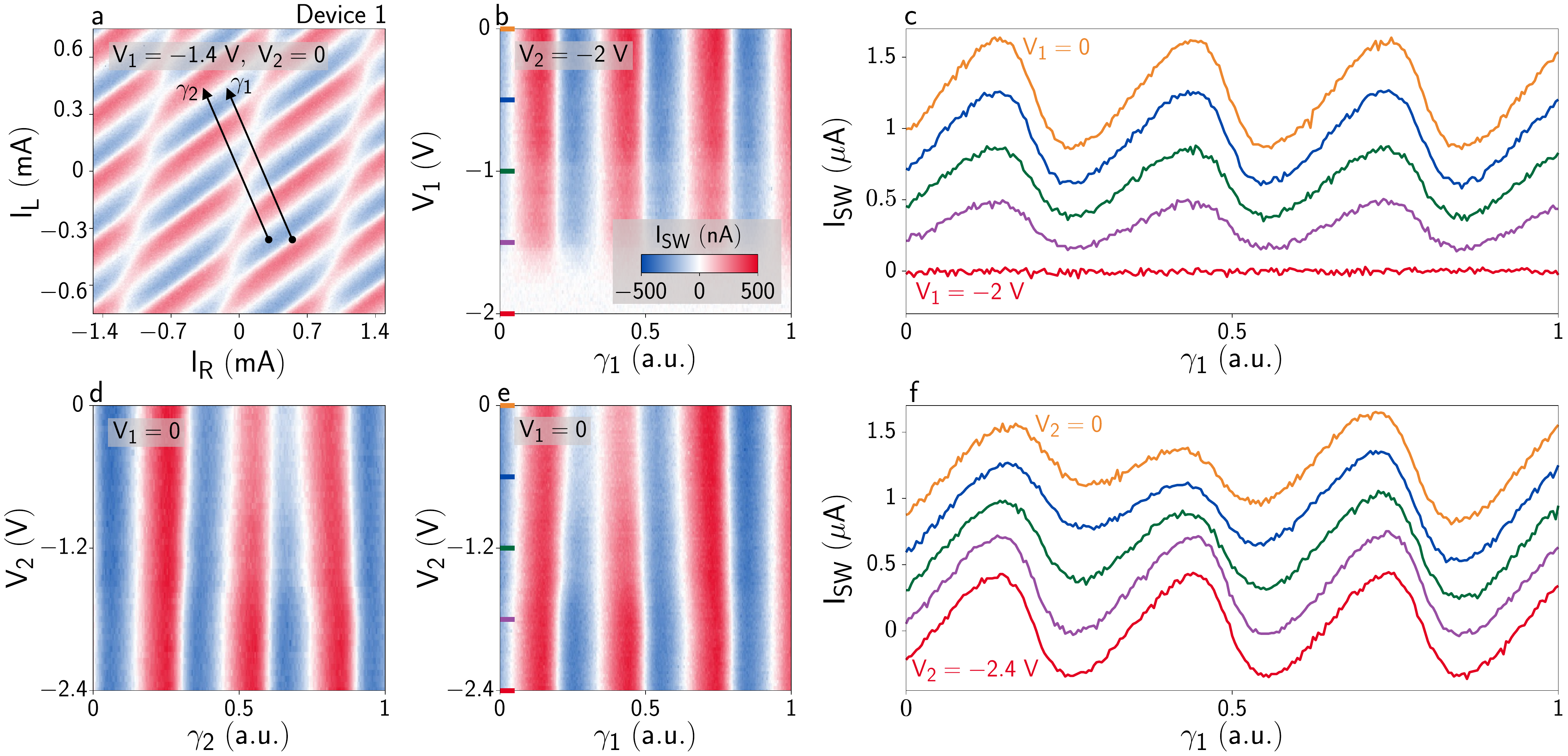}
	\caption{Gate-dependence of switching currents and nonlocal Josephson effect. (a) Switching current $\Isw$ for Device~1 as a function of $\PhiL$ and $\PhiR$, measured with $V_1=-1.4~\mathrm{V}$ and $V_2=0$. The paths $\gamma_1$ and $\gamma_2$ are shown as black arrows. (b) $\Isw$ as a function of $V_1$ and $\gamma_1$, measured with $V_2=-2~\mathrm{V}$. In this configuration, no current flows in JJ2 and $\Isw$ reflects the current--phase relation of JJ1 without hybridization. (c) Linecuts of $\Isw$ extracted from (b) for various values of $V_1$ [see markers in (b)]. (d) $\Isw$ as a function of $V_2$ and $\gamma_2$, measured with $V_1=0$. In this configuration, JJ1 was completely open. (e) As in (d), but measured along $\gamma_1$. (f) Linecuts of $\Isw$ extracted from (e) for various values of $V_2$ [see markers in (e)].}
	\label{fig3}
\end{figure*}

Figure~\ref{fig3} presents the dependence of supercurrent oscillations in Device~1 on $V_1$ and $V_2$. Panel~(a) shows supercurrent oscillations for $V_1=-1.4~\mathrm{V}$ and $V_2=0$: while the oscillation amplitude was reduced by setting $V_1$ negative, the oscillation pattern was almost identical to that of Fig.~\ref{fig2}(a), indicating that $V_1$ had negligible influence on the anomalous phase shift. The effect of gate voltages was further investigated by measuring supercurrents along the paths $\gamma_1$ and $\gamma_2$, shown as arrows in Fig.~\ref{fig3}(a). Figure~\ref{fig3}(b) shows the CPR of JJ1 measured along $\gamma_1$ with $V_2=-2~\mathrm{V}$, that is with no current flowing in JJ2. Selected linecuts are plotted in Fig.~\ref{fig3}(c). Again, we note that $V_1$ did not induce a phase shift, but simply decreased the oscillation amplitude until no current flowed in the right arm of the device. The linecuts in Fig.~\ref{fig3}(c) demonstrate that JJ1 had a forward-skewed CPR, which can be parameterized by an effective junction transmission of $\bar{\tau}=0.80$, indicating the presence of highly transmissive ABSs in JJ1 (see Supporting Information for details). Figures~\ref{fig3}(d, e) show the CPR of JJ1 along $\gamma_2$ and $\gamma_1$, respectively, measured for $V_1=0$. Selected linecuts of Fig.~\ref{fig3}(e) are shown in Fig.~\ref{fig3}(f). Both Figs.~\ref{fig3}(d) and (e) show distorted and phase-shifted supercurrent oscillations which evolved with decreasing $V_2$, saturating to a conventional forward-skewed CPR for $V_2\lesssim-2~\mathrm{V}$.

\begin{figure*}
	\includegraphics[width=\textwidth]{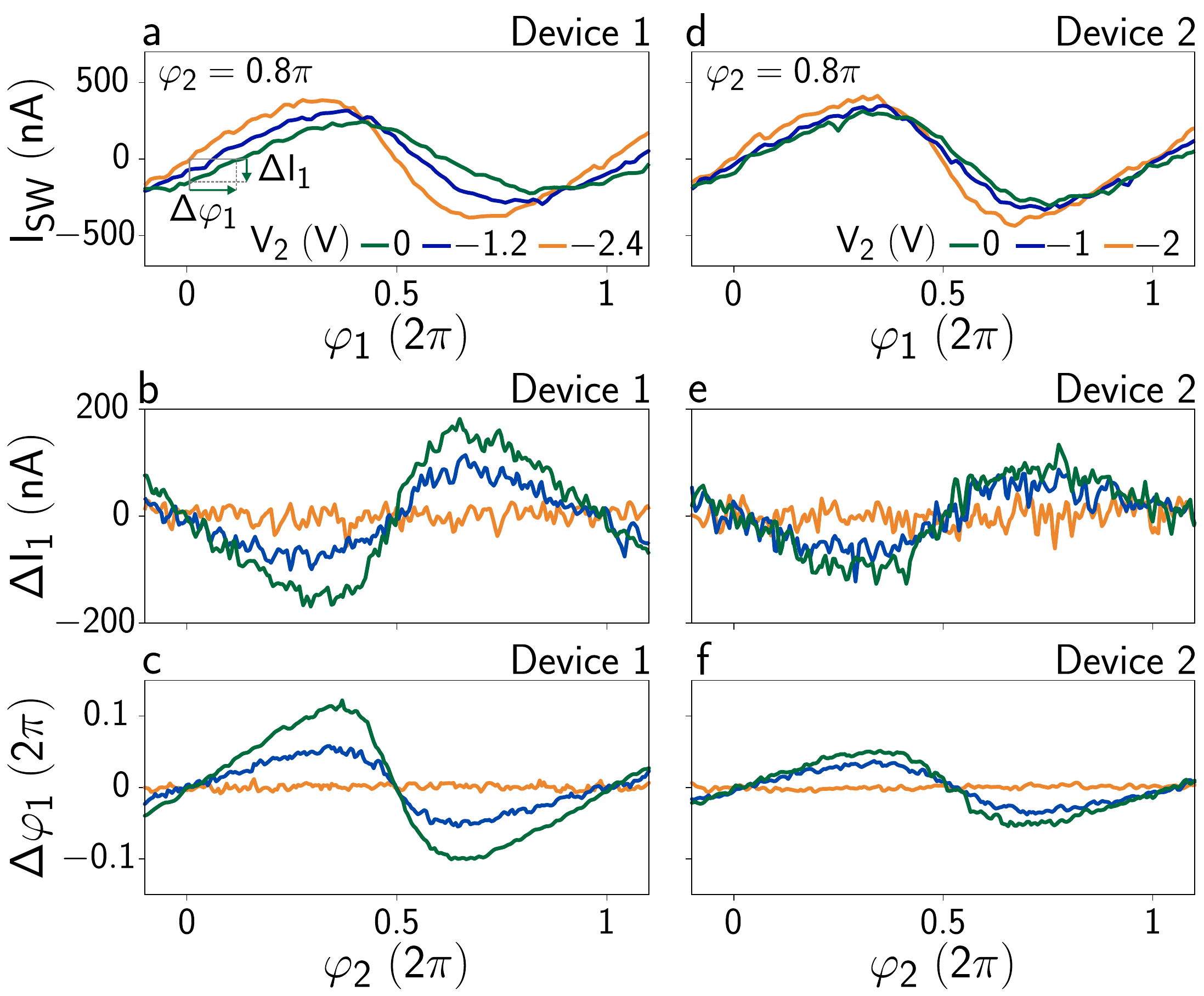}
	\caption{Anomalous supercurrent and anomalous phase shift. (a) Switching current $\Isw$ in Device~1 as a function of $\varphi_1$ measured for $\varphi_2=0.8\pi$ at three values of $V_2$. Quantities $\Delta I_1$ and $\Delta\varphi_1$ are defined. More details on data analysis required to produce this plot are presented in the Supporting Information. (b) Anomalous supercurrent $\Delta I_1$ as a function of $\varphi_2$ for three values of $V_2$ [see legend in (a)]. (c) Anomalous phase shift $\Delta\varphi_1$ as a function of $\varphi_2$ for three values of $V_2$ [see legend in (a)]. (d-f) As (a-c), but for Device~2.}
	\label{fig4}
\end{figure*}

Figure~\ref{fig4} summarizes the main results of this work. After performing an appropriate basis transformation to the data in Figs.~\ref{fig2}(a-c), it is possible to display $\Isw$ as a function of $\varphi_1$ for selected values of $\varphi_2$ [see Supporting Information for details]. For example, Fig.~\ref{fig4}(a) shows $\Isw(\varphi_1)$ for $\varphi_2=0.8\pi$, that is when the phase shift was found to be the largest. Figure~\ref{fig4}(a) also highlights the quantities $\Delta I_1$ and $\Delta\varphi_1$, which represent the anomalous supercurrent (i.e., the supercurrent at $\varphi_1=0$) and anomalous phase shift, respectively. The dependence of $\Delta I_1$ and $\Delta\varphi_1$ on $\varphi_2$ are further shown in Figs.~\ref{fig4}(b) and (c). Similar results for Device~2 are shown in Fig.~\ref{fig4}(d-f). The data presented here demonstrates coupling between the JJs, consistent with the formation of an Andreev molecule from the ABSs of JJ1 and JJ2~\cite{Pillet2019}. Time-reversal symmetry requires $I_{1}(-\varphi_{1},-\varphi_{2})=-I_{1}(\varphi_{1},\varphi_{2})$, which imposes $\Delta\varphi_{1}=0$ at $\varphi_{2}=(0,\pi)$. This condition was used to accurately account for a constant offset to $\Delta\varphi_{1}$. For $\varphi_2\neq(0,\pi)$, the CPR of JJ1 gained an anomalous phase shift $\Delta\varphi_1$ or, equivalently, an anomalous supercurrent $\Delta I_1$ at $\varphi_1=0$, controlled by both phase and gate tuning of JJ2. Both the anomalous phase shift and anomalous supercurrent were symmetric and periodic in $\varphi_2$, consistent with theoretical expectations~\cite{Pillet2019}. The phase shift $\Delta\varphi_{1}$ was also an odd function of $\varphi_{2}$, consistent with the relative interplay of dCAR and dEC processes as a function of the nonlocal phase. There was also a pronounced forward skewness to $\Delta\varphi_{1}$, which is a consequence of the nonsinusoidal CPR of both JJs. The largest $\Delta\varphi_1$ measured in Device~1 was $\pm0.22\pi$, which resulted in $\Delta I_1=\pm170~\mathrm{nA}$. These values might be increased further in devices with shorter $L$. We confirmed that coupling takes place over length scales of at least $400~\mathrm{nm}$, but significantly smaller than $4~\mathrm{\mu m}$. Such length scales are consistent with those of superconducting correlations in our devices. In the InAs quantum well, we expect a superconducting coherence length $\xi_{\mathrm{InAs}}\sim600~\mathrm{nm}$ \cite{Supplement}, compared to $\xi_{\mathrm{Al}} \sim 100~\mathrm{nm}$ of the Al film \cite{vanWoerkom2017}. These length scales are compatible with the absence of coupling in Device 3 ($L=4~\mathrm{\mu m}$), and they indicate that both the semiconductor and the superconductor might be relevant to mediate the coupling over short JJ separations. Extending our experiments to several devices with varying $L$ will make it possible to extract the typical length scale governing the nonlocal effect, thus providing a better understanding of the microscopic coupling mechanisms.

Further insights into the non-local Josephson effect are gained via the gate dependence shown in Fig.~\ref{fig3}. Data indicates that $V_1$ affected the amplitude of $\Isw$ but not $\Delta\varphi_1$, while $V_2$ tuned $\Delta\varphi_1$ without any influence on the amplitude of $\Isw$. A gate-dependent phase shift can result from tuning of the transmission and number of ABSs in each JJ, affecting the degree of hybridization between ABSs in JJ1 and JJ2. If the gate tuning of $\Delta\varphi_{1}$ was only due to a change of the ABS transmissions as a function of $V_2$, we would expect the effect to be symmetric with respect to JJ1 and JJ2, contrary to the observations. Previous work also showed that gate voltages mainly affect hybrid JJs by reducing the number of ABSs, not by changing the ABS transmission~\cite{Kjaergaard2017}. We instead speculate that the different dependence of $\Isw$ on $V_1$ and $V_2$ reflects how $\Delta\varphi_1$ is affected by the number of ABSs present in each junction. In the present case, $V_2$ controls $\Delta\varphi_1$ by tuning the number of ABSs in JJ2 available for hybridizing with the ABSs in JJ1. As JJ2 is depleted, ABSs in JJ1 progressively lose states of JJ2 to hybridize with, until a conventional CPR is restored. On the other hand, setting $V_1$ negative decreases the number of ABSs in JJ1, which directly results in a decrease of $\Isw$. However, as long as the number of ABSs in JJ2 is unchanged, the hybridization of the remaining ABSs in JJ1 is also unchanged and $\Delta\varphi_1$ remains constant.

Previous work demonstrated $\varphi_0$-junction behavior in the same material system by the combination of spin--orbit coupling and external magnetic fields~\cite{Mayer2020}. To achieve similar $\Delta\varphi_1$ as in Fig.~\ref{fig4}(c), in-plane magnetic fields of approximately $150~\mathrm{mT}$ were required. Such magnetic fields are hardly compatible with existing superconducting electronics, making the nonlocal Josephson effect mediated by Andreev molecules a particularly promising avenue to generate arbitrary phase shifts in superconducting devices. Another recent work showed a persistent $\varphi_0$-junction behavior induced by ferromagnetic elements in a JJ and field cycling, constituting a phase battery \cite{Strambini2020}; the $\varphi_0$-junction that we realized here is instead continuously and electrically tunable over short times scales, hence it could be promptly utilized as a phase source for applications in superconducting electronics and spintronics.

In conclusion, our investigation of phase-tunable JJs placed in close proximity to each other and sharing a common electrode demonstrated the formation of an Andreev molecule exhibiting nonlocal Josephson effect. This manifested as an anomalous phase shift and an anomalous supercurrent arising in one JJ, depending on phase tuning and gating of the second JJ. In the light of our results, Andreev molecules expand the available toolset of functionality in hybrid materials, also enabling new quantum manipulation schemes and coupling architectures for hybrid qubits.

Note: during the final preparation of this manuscript, a paper reporting very similar effects was posted online~\cite{Matsuo2023b}.

\section{Supporting Information}
Supporting Information includes: a detailed discussion on sample fabrication and electrical measurements; calculations of the superconducting coherence length, loop inductances and the effective junction transmission; details on current-to-flux conversion; and additional experimental results on Device~4, which was identical to Device~1.

\section{Acknowledgments}
We thank the Cleanroom Operations Team of the Binnig and Rohrer Nanotechnology Center (BRNC) for their help and support. We are grateful to H. Weisbrich and J.C. Cuevas for useful discussions. W.B.~acknowledges support from the DFG via SFB 1432 (ID 425217212) and BE 3803/14-1 (ID 467596333). W.W.~acknowledges support from the Swiss National Science Foundation (grant number 200020\_207538). F.N.~acknowledges support from the European Research Council (grant number 804273) and the Swiss National Science Foundation (grant number 200021\_201082).

\bibliography{Bibliography}

\providecommand{\latin}[1]{#1}
\makeatletter
\providecommand{\doi}
  {\begingroup\let\do\@makeother\dospecials
  \catcode`\{=1 \catcode`\}=2 \doi@aux}
\providecommand{\doi@aux}[1]{\endgroup\texttt{#1}}
\makeatother
\providecommand*\mcitethebibliography{\thebibliography}
\csname @ifundefined\endcsname{endmcitethebibliography}
  {\let\endmcitethebibliography\endthebibliography}{}
\begin{mcitethebibliography}{57}
\providecommand*\natexlab[1]{#1}
\providecommand*\mciteSetBstSublistMode[1]{}
\providecommand*\mciteSetBstMaxWidthForm[2]{}
\providecommand*\mciteBstWouldAddEndPuncttrue
  {\def\EndOfBibitem{\unskip.}}
\providecommand*\mciteBstWouldAddEndPunctfalse
  {\let\EndOfBibitem\relax}
\providecommand*\mciteSetBstMidEndSepPunct[3]{}
\providecommand*\mciteSetBstSublistLabelBeginEnd[3]{}
\providecommand*\EndOfBibitem{}
\mciteSetBstSublistMode{f}
\mciteSetBstMaxWidthForm{subitem}{(\alph{mcitesubitemcount})}
\mciteSetBstSublistLabelBeginEnd
  {\mcitemaxwidthsubitemform\space}
  {\relax}
  {\relax}

\bibitem[Krogstrup \latin{et~al.}(2015)Krogstrup, Ziino, Chang, Albrecht,
  Madsen, Johnson, Nygård, Marcus, and Jespersen]{Krogstrup2015}
Krogstrup,~P.; Ziino,~N. L.~B.; Chang,~W.; Albrecht,~S.~M.; Madsen,~M.~H.;
  Johnson,~E.; Nygård,~J.; Marcus,~C.; Jespersen,~T.~S. Epitaxy of
  semiconductor-superconductor nanowires. \emph{Nat. Mater.} \textbf{2015},
  \emph{14}, 400--406\relax
\mciteBstWouldAddEndPuncttrue
\mciteSetBstMidEndSepPunct{\mcitedefaultmidpunct}
{\mcitedefaultendpunct}{\mcitedefaultseppunct}\relax
\EndOfBibitem
\bibitem[Chang \latin{et~al.}(2015)Chang, Albrecht, Jespersen, Kuemmeth,
  Krogstrup, Nygård, and Marcus]{Chang2015}
Chang,~W.; Albrecht,~S.~M.; Jespersen,~T.~S.; Kuemmeth,~F.; Krogstrup,~P.;
  Nygård,~J.; Marcus,~C.~M. Hard gap in epitaxial semiconductor-superconductor
  nanowires. \emph{Nat. Nanotechnol.} \textbf{2015}, \emph{10}, 232--236\relax
\mciteBstWouldAddEndPuncttrue
\mciteSetBstMidEndSepPunct{\mcitedefaultmidpunct}
{\mcitedefaultendpunct}{\mcitedefaultseppunct}\relax
\EndOfBibitem
\bibitem[Shabani \latin{et~al.}(2016)Shabani, Kjaergaard, Suominen, Kim,
  Nichele, Pakrouski, Stankevic, Lutchyn, Krogstrup, Feidenhans'l, Kraemer,
  Nayak, Troyer, Marcus, and Palmstr\o{}m]{Shabani2016}
Shabani,~J.; Kjaergaard,~M.; Suominen,~H.~J.; Kim,~Y.; Nichele,~F.;
  Pakrouski,~K.; Stankevic,~T.; Lutchyn,~R.~M.; Krogstrup,~P.;
  Feidenhans'l,~R.; Kraemer,~S.; Nayak,~C.; Troyer,~M.; Marcus,~C.~M.;
  Palmstr\o{}m,~C.~J. Two-dimensional epitaxial superconductor-semiconductor
  heterostructures: A platform for topological superconducting networks.
  \emph{Phys. Rev. B} \textbf{2016}, \emph{93}, 155402\relax
\mciteBstWouldAddEndPuncttrue
\mciteSetBstMidEndSepPunct{\mcitedefaultmidpunct}
{\mcitedefaultendpunct}{\mcitedefaultseppunct}\relax
\EndOfBibitem
\bibitem[Andreev(1964)]{Andreev1964}
Andreev,~A.~F. Thermal conductivity of the intermediate state of
  superconductors. \emph{Sov. Phys. JETP} \textbf{1964}, \emph{19},
  1228--1231\relax
\mciteBstWouldAddEndPuncttrue
\mciteSetBstMidEndSepPunct{\mcitedefaultmidpunct}
{\mcitedefaultendpunct}{\mcitedefaultseppunct}\relax
\EndOfBibitem
\bibitem[Beenakker and van Houten(1991)Beenakker, and van
  Houten]{Beenakker1991}
Beenakker,~C. W.~J.; van Houten,~H. {J}osephson current through a
  superconducting quantum point contact shorter than the coherence length.
  \emph{Phys. Rev. Lett.} \textbf{1991}, \emph{66}, 3056--3059\relax
\mciteBstWouldAddEndPuncttrue
\mciteSetBstMidEndSepPunct{\mcitedefaultmidpunct}
{\mcitedefaultendpunct}{\mcitedefaultseppunct}\relax
\EndOfBibitem
\bibitem[Furusaki and Tsukada(1991)Furusaki, and Tsukada]{Furusaki1991}
Furusaki,~A.; Tsukada,~M. Current-carrying states in {J}osephson junctions.
  \emph{Phys. Rev. B} \textbf{1991}, \emph{43}, 10164--10169\relax
\mciteBstWouldAddEndPuncttrue
\mciteSetBstMidEndSepPunct{\mcitedefaultmidpunct}
{\mcitedefaultendpunct}{\mcitedefaultseppunct}\relax
\EndOfBibitem
\bibitem[Pillet \latin{et~al.}(2010)Pillet, Quay, Morfin, Bena, Yeyati, and
  Joyez]{Pillet2010}
Pillet,~J.-D.; Quay,~C. H.~L.; Morfin,~P.; Bena,~C.; Yeyati,~A.~L.; Joyez,~P.
  Andreev bound states in supercurrent-carrying carbon nanotubes revealed.
  \emph{Nat. Phys.} \textbf{2010}, \emph{6}, 965--969\relax
\mciteBstWouldAddEndPuncttrue
\mciteSetBstMidEndSepPunct{\mcitedefaultmidpunct}
{\mcitedefaultendpunct}{\mcitedefaultseppunct}\relax
\EndOfBibitem
\bibitem[Bretheau \latin{et~al.}(2013)Bretheau, Girit, Urbina, Esteve, and
  Pothier]{Bretheau2013b}
Bretheau,~L.; Girit,~{\c{C}}.~{\"O}.; Urbina,~C.; Esteve,~D.; Pothier,~H.
  Supercurrent Spectroscopy of {A}ndreev States. \emph{Phys. Rev. X}
  \textbf{2013}, \emph{3}, 041034\relax
\mciteBstWouldAddEndPuncttrue
\mciteSetBstMidEndSepPunct{\mcitedefaultmidpunct}
{\mcitedefaultendpunct}{\mcitedefaultseppunct}\relax
\EndOfBibitem
\bibitem[Tosi \latin{et~al.}(2019)Tosi, Metzger, Goffman, Urbina, Pothier,
  Park, Yeyati, Nyg\aa{}rd, and Krogstrup]{Tosi2019}
Tosi,~L.; Metzger,~C.; Goffman,~M.~F.; Urbina,~C.; Pothier,~H.; Park,~S.;
  Yeyati,~A.~L.; Nyg\aa{}rd,~J.; Krogstrup,~P. Spin--orbit splitting of
  {A}ndreev states revealed by microwave spectroscopy. \emph{Phys. Rev. X}
  \textbf{2019}, \emph{9}, 011010\relax
\mciteBstWouldAddEndPuncttrue
\mciteSetBstMidEndSepPunct{\mcitedefaultmidpunct}
{\mcitedefaultendpunct}{\mcitedefaultseppunct}\relax
\EndOfBibitem
\bibitem[Nichele \latin{et~al.}(2020)Nichele, Portol\'es, Fornieri, Whiticar,
  Drachmann, Gronin, Wang, Gardner, Thomas, Hatke, Manfra, and
  Marcus]{Nichele2020}
Nichele,~F.; Portol\'es,~E.; Fornieri,~A.; Whiticar,~A.~M.; Drachmann,~A.
  C.~C.; Gronin,~S.; Wang,~T.; Gardner,~G.~C.; Thomas,~C.; Hatke,~A.~T.;
  Manfra,~M.~J.; Marcus,~C.~M. Relating {A}ndreev bound states and
  supercurrents in hybrid {J}osephson junctions. \emph{Phys. Rev. Lett.}
  \textbf{2020}, \emph{124}, 226801\relax
\mciteBstWouldAddEndPuncttrue
\mciteSetBstMidEndSepPunct{\mcitedefaultmidpunct}
{\mcitedefaultendpunct}{\mcitedefaultseppunct}\relax
\EndOfBibitem
\bibitem[Doh \latin{et~al.}(2005)Doh, van Dam, Roest, Bakkers, Kouwenhoven, and
  Franceschi]{Doh2005}
Doh,~Y.-J.; van Dam,~J.~A.; Roest,~A.~L.; Bakkers,~E. P. A.~M.;
  Kouwenhoven,~L.~P.; Franceschi,~S.~D. Tunable supercurrent through
  semiconductor nanowires. \emph{Science} \textbf{2005}, \emph{309},
  272--275\relax
\mciteBstWouldAddEndPuncttrue
\mciteSetBstMidEndSepPunct{\mcitedefaultmidpunct}
{\mcitedefaultendpunct}{\mcitedefaultseppunct}\relax
\EndOfBibitem
\bibitem[Xiang \latin{et~al.}(2006)Xiang, Vidan, Tinkham, Westervelt, and
  Lieber]{Xiang2006}
Xiang,~J.; Vidan,~A.; Tinkham,~M.; Westervelt,~R.~M.; Lieber,~C.~M. Ge/Si
  nanowire mesoscopic Josephson junctions. \emph{Nat. Nanotechnol.}
  \textbf{2006}, \emph{1}, 208--213\relax
\mciteBstWouldAddEndPuncttrue
\mciteSetBstMidEndSepPunct{\mcitedefaultmidpunct}
{\mcitedefaultendpunct}{\mcitedefaultseppunct}\relax
\EndOfBibitem
\bibitem[Gheewala(1980)]{Gheewala1980}
Gheewala,~T. Josephson-logic devices and circuits. \emph{IEEE Trans. Electron
  Devices} \textbf{1980}, \emph{27}, 1857--1869\relax
\mciteBstWouldAddEndPuncttrue
\mciteSetBstMidEndSepPunct{\mcitedefaultmidpunct}
{\mcitedefaultendpunct}{\mcitedefaultseppunct}\relax
\EndOfBibitem
\bibitem[Clark \latin{et~al.}(1980)Clark, Prance, and Grassie]{Clark1980}
Clark,~T.~D.; Prance,~R.~J.; Grassie,~A. D.~C. Feasibility of hybrid Josephson
  field effect transistors. \emph{J. Appl. Phys.} \textbf{1980}, \emph{51},
  2736\relax
\mciteBstWouldAddEndPuncttrue
\mciteSetBstMidEndSepPunct{\mcitedefaultmidpunct}
{\mcitedefaultendpunct}{\mcitedefaultseppunct}\relax
\EndOfBibitem
\bibitem[Kleinsasser and Jackson(1989)Kleinsasser, and
  Jackson]{Kleinsasser1989}
Kleinsasser,~A.; Jackson,~T. Prospects for proximity effect superconducting
  {FETs}. \emph{IEEE Trans. Magn.} \textbf{1989}, \emph{25}, 1274--1277\relax
\mciteBstWouldAddEndPuncttrue
\mciteSetBstMidEndSepPunct{\mcitedefaultmidpunct}
{\mcitedefaultendpunct}{\mcitedefaultseppunct}\relax
\EndOfBibitem
\bibitem[Wen \latin{et~al.}(2019)Wen, Shabani, and Tutuc]{Wen2019}
Wen,~F.; Shabani,~J.; Tutuc,~E. {J}osephson junction field-effect transistors
  for boolean logic cryogenic applications. \emph{IEEE Trans. Electron Devices}
  \textbf{2019}, \emph{66}, 5367--5374\relax
\mciteBstWouldAddEndPuncttrue
\mciteSetBstMidEndSepPunct{\mcitedefaultmidpunct}
{\mcitedefaultendpunct}{\mcitedefaultseppunct}\relax
\EndOfBibitem
\bibitem[Larsen \latin{et~al.}(2015)Larsen, Petersson, Kuemmeth, Jespersen,
  Krogstrup, Nyg\aa{}rd, and Marcus]{Larsen2015}
Larsen,~T.~W.; Petersson,~K.~D.; Kuemmeth,~F.; Jespersen,~T.~S.; Krogstrup,~P.;
  Nyg\aa{}rd,~J.; Marcus,~C.~M. Semiconductor-nanowire-based superconducting
  qubit. \emph{Phys. Rev. Lett.} \textbf{2015}, \emph{115}, 127001\relax
\mciteBstWouldAddEndPuncttrue
\mciteSetBstMidEndSepPunct{\mcitedefaultmidpunct}
{\mcitedefaultendpunct}{\mcitedefaultseppunct}\relax
\EndOfBibitem
\bibitem[de~Lange \latin{et~al.}(2015)de~Lange, van Heck, Bruno, van Woerkom,
  Geresdi, Plissard, Bakkers, Akhmerov, and DiCarlo]{deLange2015}
de~Lange,~G.; van Heck,~B.; Bruno,~A.; van Woerkom,~D.~J.; Geresdi,~A.;
  Plissard,~S.~R.; Bakkers,~E. P. A.~M.; Akhmerov,~A.~R.; DiCarlo,~L.
  Realization of microwave quantum circuits using hybrid
  superconducting-semiconducting nanowire {J}osephson elements. \emph{Phys.
  Rev. Lett.} \textbf{2015}, \emph{115}, 127002\relax
\mciteBstWouldAddEndPuncttrue
\mciteSetBstMidEndSepPunct{\mcitedefaultmidpunct}
{\mcitedefaultendpunct}{\mcitedefaultseppunct}\relax
\EndOfBibitem
\bibitem[Casparis \latin{et~al.}(2018)Casparis, Connolly, Kjaergaard, Pearson,
  Kringh{\o}j, Larsen, Kuemmeth, Wang, Thomas, Gronin, Gardner, Manfra, Marcus,
  and Petersson]{Casparis2018}
Casparis,~L.; Connolly,~M.~R.; Kjaergaard,~M.; Pearson,~N.~J.; Kringh{\o}j,~A.;
  Larsen,~T.~W.; Kuemmeth,~F.; Wang,~T.; Thomas,~C.; Gronin,~S.;
  Gardner,~G.~C.; Manfra,~M.~J.; Marcus,~C.~M.; Petersson,~K.~D.
  {Superconducting gatemon qubit based on a proximitized two-dimensional
  electron gas}. \emph{Nat. Nanotechnol.} \textbf{2018}, \emph{13},
  915--919\relax
\mciteBstWouldAddEndPuncttrue
\mciteSetBstMidEndSepPunct{\mcitedefaultmidpunct}
{\mcitedefaultendpunct}{\mcitedefaultseppunct}\relax
\EndOfBibitem
\bibitem[Pita-Vidal \latin{et~al.}(2020)Pita-Vidal, Bargerbos, Yang, van
  Woerkom, Pfaff, Haider, Krogstrup, Kouwenhoven, de~Lange, and
  Kou]{PitaVidal2020}
Pita-Vidal,~M.; Bargerbos,~A.; Yang,~C.-K.; van Woerkom,~D.~J.; Pfaff,~W.;
  Haider,~N.; Krogstrup,~P.; Kouwenhoven,~L.~P.; de~Lange,~G.; Kou,~A.
  Gate-tunable field-compatible fluxonium. \emph{Phys. Rev. Appl.}
  \textbf{2020}, \emph{14}, 064038\relax
\mciteBstWouldAddEndPuncttrue
\mciteSetBstMidEndSepPunct{\mcitedefaultmidpunct}
{\mcitedefaultendpunct}{\mcitedefaultseppunct}\relax
\EndOfBibitem
\bibitem[Casparis \latin{et~al.}(2019)Casparis, Pearson, Kringh\o{}j, Larsen,
  Kuemmeth, Nyg\aa{}rd, Krogstrup, Petersson, and Marcus]{Casparis2019}
Casparis,~L.; Pearson,~N.~J.; Kringh\o{}j,~A.; Larsen,~T.~W.; Kuemmeth,~F.;
  Nyg\aa{}rd,~J.; Krogstrup,~P.; Petersson,~K.~D.; Marcus,~C.~M.
  Voltage-controlled superconducting quantum bus. \emph{Phys. Rev. B}
  \textbf{2019}, \emph{99}, 085434\relax
\mciteBstWouldAddEndPuncttrue
\mciteSetBstMidEndSepPunct{\mcitedefaultmidpunct}
{\mcitedefaultendpunct}{\mcitedefaultseppunct}\relax
\EndOfBibitem
\bibitem[Sardashti \latin{et~al.}(2020)Sardashti, Dartiailh, Yuan, Hart,
  Gumann, and Shabani]{Sardashti2020}
Sardashti,~K.; Dartiailh,~M.~C.; Yuan,~J.; Hart,~S.; Gumann,~P.; Shabani,~J.
  Voltage-tunable superconducting resonators: a platform for random access
  quantum memory. \emph{IEEE Trans. Quantum Eng.} \textbf{2020}, \emph{1},
  1--7\relax
\mciteBstWouldAddEndPuncttrue
\mciteSetBstMidEndSepPunct{\mcitedefaultmidpunct}
{\mcitedefaultendpunct}{\mcitedefaultseppunct}\relax
\EndOfBibitem
\bibitem[Butseraen \latin{et~al.}(2022)Butseraen, Ranadive, Aparicio,
  Rafsanjani~Amin, Juyal, Esposito, Watanabe, Taniguchi, Roch, Lefloch, and
  Renard]{Butseraen2022}
Butseraen,~G.; Ranadive,~A.; Aparicio,~N.; Rafsanjani~Amin,~K.; Juyal,~A.;
  Esposito,~M.; Watanabe,~K.; Taniguchi,~T.; Roch,~N.; Lefloch,~F.; Renard,~J.
  A gate-tunable graphene Josephson parametric amplifier. \emph{Nat.
  Nanotechnol.} \textbf{2022}, \emph{17}, 1153--1158\relax
\mciteBstWouldAddEndPuncttrue
\mciteSetBstMidEndSepPunct{\mcitedefaultmidpunct}
{\mcitedefaultendpunct}{\mcitedefaultseppunct}\relax
\EndOfBibitem
\bibitem[Sarkar \latin{et~al.}(2022)Sarkar, Salunkhe, Mandal, Ghatak,
  Marchawala, Das, Watanabe, Taniguchi, Vijay, and Deshmukh]{Sarkar2022}
Sarkar,~J.; Salunkhe,~K.~V.; Mandal,~S.; Ghatak,~S.; Marchawala,~A.~H.;
  Das,~I.; Watanabe,~K.; Taniguchi,~T.; Vijay,~R.; Deshmukh,~M.~M.
  Quantum-noise-limited microwave amplification using a graphene Josephson
  junction. \emph{Nat. Nanotechnol} \textbf{2022}, \emph{17}, 1147--1152\relax
\mciteBstWouldAddEndPuncttrue
\mciteSetBstMidEndSepPunct{\mcitedefaultmidpunct}
{\mcitedefaultendpunct}{\mcitedefaultseppunct}\relax
\EndOfBibitem
\bibitem[Phan \latin{et~al.}(2022)Phan, Falthansl-Scheinecker, Mishra,
  Strickland, Langone, Shabani, and Higginbotham]{Phan2023}
Phan,~D.; Falthansl-Scheinecker,~P.; Mishra,~U.; Strickland,~W.~M.;
  Langone,~D.; Shabani,~J.; Higginbotham,~A.~P. Gate-tunable,
  superconductor-semiconductor parametric amplifier. arXiv:2206.05746,
  2022\relax
\mciteBstWouldAddEndPuncttrue
\mciteSetBstMidEndSepPunct{\mcitedefaultmidpunct}
{\mcitedefaultendpunct}{\mcitedefaultseppunct}\relax
\EndOfBibitem
\bibitem[Baumgartner \latin{et~al.}(2022)Baumgartner, Fuchs, Costa, Reinhardt,
  Gronin, Gardner, Lindemann, Manfra, {Faria Junior}, Kochan, Fabian, Paradiso,
  and Strunk]{Baumgartner2022}
Baumgartner,~C.; Fuchs,~L.; Costa,~A.; Reinhardt,~S.; Gronin,~S.;
  Gardner,~G.~C.; Lindemann,~T.; Manfra,~M.~J.; {Faria Junior},~P.~E.;
  Kochan,~D.; Fabian,~J.; Paradiso,~N.; Strunk,~C. {Supercurrent rectification
  and magnetochiral effects in symmetric {J}osephson junctions}. \emph{Nat.
  Nanotechnol.} \textbf{2022}, \emph{17}, 39--44\relax
\mciteBstWouldAddEndPuncttrue
\mciteSetBstMidEndSepPunct{\mcitedefaultmidpunct}
{\mcitedefaultendpunct}{\mcitedefaultseppunct}\relax
\EndOfBibitem
\bibitem[Turini \latin{et~al.}(2022)Turini, Salimian, Carrega, Iorio,
  Strambini, Giazotto, Zannier, Sorba, and Heun]{Turini2022}
Turini,~B.; Salimian,~S.; Carrega,~M.; Iorio,~A.; Strambini,~E.; Giazotto,~F.;
  Zannier,~V.; Sorba,~L.; Heun,~S. Josephson diode effect in high-mobility InSb
  nanoflags. \emph{Nano Lett.} \textbf{2022}, \emph{22}, 8502--8508\relax
\mciteBstWouldAddEndPuncttrue
\mciteSetBstMidEndSepPunct{\mcitedefaultmidpunct}
{\mcitedefaultendpunct}{\mcitedefaultseppunct}\relax
\EndOfBibitem
\bibitem[Gupta \latin{et~al.}(2023)Gupta, Graziano, Pendharkar, Dong, Dempsey,
  Palmstrøm, and Pribiag]{Gupta2023}
Gupta,~M.; Graziano,~G.~V.; Pendharkar,~M.; Dong,~J.~T.; Dempsey,~C.~P.;
  Palmstrøm,~C.; Pribiag,~V.~S. Gate-tunable superconducting diode effect in a
  three-terminal Josephson device. \emph{Nat. Commun.} \textbf{2023},
  \emph{14}, 3078\relax
\mciteBstWouldAddEndPuncttrue
\mciteSetBstMidEndSepPunct{\mcitedefaultmidpunct}
{\mcitedefaultendpunct}{\mcitedefaultseppunct}\relax
\EndOfBibitem
\bibitem[Matsuo \latin{et~al.}(2023)Matsuo, Imoto, Yokoyama, Sato, Lindemann,
  Gronin, Gardner, Manfra, and Tarucha]{Matsuo2023c}
Matsuo,~S.; Imoto,~T.; Yokoyama,~T.; Sato,~Y.; Lindemann,~T.; Gronin,~S.;
  Gardner,~G.~C.; Manfra,~M.~J.; Tarucha,~S. Josephson diode effect derived
  from short-range coherent coupling. arXiv:2305.07923, 2023\relax
\mciteBstWouldAddEndPuncttrue
\mciteSetBstMidEndSepPunct{\mcitedefaultmidpunct}
{\mcitedefaultendpunct}{\mcitedefaultseppunct}\relax
\EndOfBibitem
\bibitem[Buzdin and Koshelev(2003)Buzdin, and Koshelev]{Buzdin2003}
Buzdin,~A.; Koshelev,~A.~E. Periodic alternating 0- and
  $\ensuremath{\pi}$-junction structures as realization of
  $\ensuremath{\varphi}$-{J}osephson junctions. \emph{Phys. Rev. B}
  \textbf{2003}, \emph{67}, 220504\relax
\mciteBstWouldAddEndPuncttrue
\mciteSetBstMidEndSepPunct{\mcitedefaultmidpunct}
{\mcitedefaultendpunct}{\mcitedefaultseppunct}\relax
\EndOfBibitem
\bibitem[Buzdin(2008)]{Buzdin2008}
Buzdin,~A. Direct coupling between magnetism and superconducting current in the
  {J}osephson ${\ensuremath{\varphi}}_{0}$ junction. \emph{Phys. Rev. Lett.}
  \textbf{2008}, \emph{101}, 107005\relax
\mciteBstWouldAddEndPuncttrue
\mciteSetBstMidEndSepPunct{\mcitedefaultmidpunct}
{\mcitedefaultendpunct}{\mcitedefaultseppunct}\relax
\EndOfBibitem
\bibitem[Yokoyama \latin{et~al.}(2014)Yokoyama, Eto, and Nazarov]{Yokoyama2014}
Yokoyama,~T.; Eto,~M.; Nazarov,~Y.~V. Anomalous {J}osephson effect induced by
  spin-orbit interaction and Zeeman effect in semiconductor nanowires.
  \emph{Phys. Rev. B} \textbf{2014}, \emph{89}, 195407\relax
\mciteBstWouldAddEndPuncttrue
\mciteSetBstMidEndSepPunct{\mcitedefaultmidpunct}
{\mcitedefaultendpunct}{\mcitedefaultseppunct}\relax
\EndOfBibitem
\bibitem[Bergeret and Tokatly(2015)Bergeret, and Tokatly]{Bergeret2015}
Bergeret,~F.~S.; Tokatly,~I.~V. Theory of diffusive $\varphi$-0 Josephson
  junctions in the presence of spin-orbit coupling. \emph{EPL} \textbf{2015},
  \emph{110}, 57005\relax
\mciteBstWouldAddEndPuncttrue
\mciteSetBstMidEndSepPunct{\mcitedefaultmidpunct}
{\mcitedefaultendpunct}{\mcitedefaultseppunct}\relax
\EndOfBibitem
\bibitem[Szombati \latin{et~al.}(2016)Szombati, Nadj-Perge, Car, Plissard,
  Bakkers, and Kouwenhoven]{Szombati2016}
Szombati,~D.~B.; Nadj-Perge,~S.; Car,~D.; Plissard,~S.~R.; Bakkers,~E. P.
  A.~M.; Kouwenhoven,~L.~P. Josephson $\varphi_{0}$-junction in nanowire
  quantum dots. \emph{Nat. Phys.} \textbf{2016}, \emph{12}, 568--572\relax
\mciteBstWouldAddEndPuncttrue
\mciteSetBstMidEndSepPunct{\mcitedefaultmidpunct}
{\mcitedefaultendpunct}{\mcitedefaultseppunct}\relax
\EndOfBibitem
\bibitem[Hart \latin{et~al.}(2017)Hart, Ren, Kosowsky, Ben-Shach, Leubner,
  Br{\"{u}}ne, Buhmann, Molenkamp, Halperin, and Yacoby]{Hart2017}
Hart,~S.; Ren,~H.; Kosowsky,~M.; Ben-Shach,~G.; Leubner,~P.; Br{\"{u}}ne,~C.;
  Buhmann,~H.; Molenkamp,~L.~W.; Halperin,~B.~I.; Yacoby,~A. {Controlled finite
  momentum pairing and spatially varying order parameter in proximitized
  {H}g{T}e quantum wells}. \emph{Nat. Phys.} \textbf{2017}, \emph{13},
  87--93\relax
\mciteBstWouldAddEndPuncttrue
\mciteSetBstMidEndSepPunct{\mcitedefaultmidpunct}
{\mcitedefaultendpunct}{\mcitedefaultseppunct}\relax
\EndOfBibitem
\bibitem[Assouline \latin{et~al.}(2019)Assouline, Feuillet-Palma, Bergeal,
  Zhang, Mottaghizadeh, Zimmers, Lhuillier, Eddrie, Atkinson, Aprili, and
  Aubin]{Assouline2019}
Assouline,~A.; Feuillet-Palma,~C.; Bergeal,~N.; Zhang,~T.; Mottaghizadeh,~A.;
  Zimmers,~A.; Lhuillier,~E.; Eddrie,~M.; Atkinson,~P.; Aprili,~M.; Aubin,~H.
  {Spin--orbit induced phase-shift in {B}i2{S}e3 {J}osephson junctions}.
  \emph{Nat. Commun.} \textbf{2019}, \emph{10}, 126\relax
\mciteBstWouldAddEndPuncttrue
\mciteSetBstMidEndSepPunct{\mcitedefaultmidpunct}
{\mcitedefaultendpunct}{\mcitedefaultseppunct}\relax
\EndOfBibitem
\bibitem[Mayer \latin{et~al.}(2020)Mayer, Dartiailh, Yuan, Wickramasinghe,
  Rossi, and Shabani]{Mayer2020}
Mayer,~W.; Dartiailh,~M.~C.; Yuan,~J.; Wickramasinghe,~K.~S.; Rossi,~E.;
  Shabani,~J. {Gate controlled anomalous phase shift in Al/InAs Josephson
  junctions}. \emph{Nat. Commun.} \textbf{2020}, \emph{11}, 212\relax
\mciteBstWouldAddEndPuncttrue
\mciteSetBstMidEndSepPunct{\mcitedefaultmidpunct}
{\mcitedefaultendpunct}{\mcitedefaultseppunct}\relax
\EndOfBibitem
\bibitem[Strambini \latin{et~al.}(2020)Strambini, Iorio, Durante, Citro,
  Sanz-Fernández, Guarcello, Tokatly, Braggio, Rocci, Ligato, Zannier, Sorba,
  Bergeret, and Giazotto]{Strambini2020}
Strambini,~E.; Iorio,~A.; Durante,~O.; Citro,~R.; Sanz-Fernández,~C.;
  Guarcello,~C.; Tokatly,~I.~V.; Braggio,~A.; Rocci,~M.; Ligato,~N.;
  Zannier,~V.; Sorba,~L.; Bergeret,~F.~S.; Giazotto,~F. A Josephson phase
  battery. \emph{Nat. Nanotechnol.} \textbf{2020}, \emph{15}, 656--660\relax
\mciteBstWouldAddEndPuncttrue
\mciteSetBstMidEndSepPunct{\mcitedefaultmidpunct}
{\mcitedefaultendpunct}{\mcitedefaultseppunct}\relax
\EndOfBibitem
\bibitem[Linder and Robinson(2015)Linder, and Robinson]{Linder2015}
Linder,~J.; Robinson,~J. W.~A. Superconducting spintronics. \emph{Nat. Phys.}
  \textbf{2015}, \emph{11}, 307--315\relax
\mciteBstWouldAddEndPuncttrue
\mciteSetBstMidEndSepPunct{\mcitedefaultmidpunct}
{\mcitedefaultendpunct}{\mcitedefaultseppunct}\relax
\EndOfBibitem
\bibitem[Pillet \latin{et~al.}(2019)Pillet, Benzoni, Griesmar, Smirr, and
  Girit]{Pillet2019}
Pillet,~J.-D.; Benzoni,~V.; Griesmar,~J.; Smirr,~J.-L.; Girit,~{\c{C}}.~O.
  Nonlocal {J}osephson effect in {A}ndreev molecules. \emph{Nano Lett.}
  \textbf{2019}, \emph{19}, 7138--7143\relax
\mciteBstWouldAddEndPuncttrue
\mciteSetBstMidEndSepPunct{\mcitedefaultmidpunct}
{\mcitedefaultendpunct}{\mcitedefaultseppunct}\relax
\EndOfBibitem
\bibitem[Kornich \latin{et~al.}(2019)Kornich, Barakov, and
  Nazarov]{Kornich2019}
Kornich,~V.; Barakov,~H.~S.; Nazarov,~Y.~V. Fine energy splitting of
  overlapping {A}ndreev bound states in multiterminal superconducting
  nanostructures. \emph{Phys. Rev. Res.} \textbf{2019}, \emph{1}, 033004\relax
\mciteBstWouldAddEndPuncttrue
\mciteSetBstMidEndSepPunct{\mcitedefaultmidpunct}
{\mcitedefaultendpunct}{\mcitedefaultseppunct}\relax
\EndOfBibitem
\bibitem[Pillet \latin{et~al.}(2020)Pillet, Benzoni, Griesmar, Smirr, and
  Girit]{Pillet2020}
Pillet,~J.-D.; Benzoni,~V.; Griesmar,~J.; Smirr,~J.-L.; Girit,~{\c{C}}.
  Scattering description of {A}ndreev molecules. \emph{{SciPost} Phys.~Core}
  \textbf{2020}, \emph{2}, 009\relax
\mciteBstWouldAddEndPuncttrue
\mciteSetBstMidEndSepPunct{\mcitedefaultmidpunct}
{\mcitedefaultendpunct}{\mcitedefaultseppunct}\relax
\EndOfBibitem
\bibitem[Kornich \latin{et~al.}(2020)Kornich, Barakov, and
  Nazarov]{Kornich2020}
Kornich,~V.; Barakov,~H.~S.; Nazarov,~Y.~V. Overlapping {A}ndreev states in
  semiconducting nanowires: {C}ompetition of one-dimensional and
  three-dimensional propagation. \emph{Phys. Rev. B} \textbf{2020}, \emph{101},
  195430\relax
\mciteBstWouldAddEndPuncttrue
\mciteSetBstMidEndSepPunct{\mcitedefaultmidpunct}
{\mcitedefaultendpunct}{\mcitedefaultseppunct}\relax
\EndOfBibitem
\bibitem[Su \latin{et~al.}(2017)Su, Tacla, Hocevar, Car, Plissard, Bakkers,
  Daley, Pekker, and Frolov]{Su2017}
Su,~Z.; Tacla,~A.~B.; Hocevar,~M.; Car,~D.; Plissard,~S.~R.; Bakkers,~E. P.
  A.~M.; Daley,~A.~J.; Pekker,~D.; Frolov,~S.~M. Andreev molecules in
  semiconductor nanowire double quantum dots. \emph{Nat. Commun.}
  \textbf{2017}, \emph{8}, 585\relax
\mciteBstWouldAddEndPuncttrue
\mciteSetBstMidEndSepPunct{\mcitedefaultmidpunct}
{\mcitedefaultendpunct}{\mcitedefaultseppunct}\relax
\EndOfBibitem
\bibitem[K\"{u}rt\"{o}ssy \latin{et~al.}(2021)K\"{u}rt\"{o}ssy, Scher\"{u}bl,
  F\"{u}l\"{o}p, Luk{\'{a}}cs, Kanne, Nyg{\aa}rd, Makk, and
  Csonka]{Kurtossy2021}
K\"{u}rt\"{o}ssy,~O.; Scher\"{u}bl,~Z.; F\"{u}l\"{o}p,~G.; Luk{\'{a}}cs,~I.~E.;
  Kanne,~T.; Nyg{\aa}rd,~J.; Makk,~P.; Csonka,~S. Andreev molecule in parallel
  {InAs} nanowires. \emph{Nano Lett.} \textbf{2021}, \emph{21},
  7929--7937\relax
\mciteBstWouldAddEndPuncttrue
\mciteSetBstMidEndSepPunct{\mcitedefaultmidpunct}
{\mcitedefaultendpunct}{\mcitedefaultseppunct}\relax
\EndOfBibitem
\bibitem[Dvir \latin{et~al.}(2023)Dvir, Wang, van Loo, Liu, Mazur, Bordin, ten
  Haaf, Wang, van Driel, Zatelli, Li, Malinowski, Gazibegovic, Badawy, Bakkers,
  Wimmer, and Kouwenhoven]{Dvir2023}
Dvir,~T. \latin{et~al.}  Realization of a minimal Kitaev chain in coupled
  quantum dots. \emph{Nature} \textbf{2023}, \emph{614}, 445--450\relax
\mciteBstWouldAddEndPuncttrue
\mciteSetBstMidEndSepPunct{\mcitedefaultmidpunct}
{\mcitedefaultendpunct}{\mcitedefaultseppunct}\relax
\EndOfBibitem
\bibitem[Coraiola \latin{et~al.}(2023)Coraiola, Haxell, Sabonis, Weisbrich,
  Svetogorov, Hinderling, ten Kate, Cheah, Krizek, Schott, Wegscheider, Cuevas,
  Belzig, and Nichele]{Coraiola2023}
Coraiola,~M.; Haxell,~D.~Z.; Sabonis,~D.; Weisbrich,~H.; Svetogorov,~A.~E.;
  Hinderling,~M.; ten Kate,~S.~C.; Cheah,~E.; Krizek,~F.; Schott,~R.;
  Wegscheider,~W.; Cuevas,~J.~C.; Belzig,~W.; Nichele,~F. Hybridisation of
  {A}ndreev bound states in three-terminal {J}osephson junctions.
  arXiv:2302.14535, 2023\relax
\mciteBstWouldAddEndPuncttrue
\mciteSetBstMidEndSepPunct{\mcitedefaultmidpunct}
{\mcitedefaultendpunct}{\mcitedefaultseppunct}\relax
\EndOfBibitem
\bibitem[Matsuo \latin{et~al.}(2022)Matsuo, Lee, Chang, Sato, Ueda,
  Palmstr{\o}m, and Tarucha]{Matsuo2022}
Matsuo,~S.; Lee,~J.~S.; Chang,~C.-Y.; Sato,~Y.; Ueda,~K.; Palmstr{\o}m,~C.~J.;
  Tarucha,~S. Observation of nonlocal {J}osephson effect on double {InAs}
  nanowires. \emph{Commun. Phys.} \textbf{2022}, \emph{5}, 221\relax
\mciteBstWouldAddEndPuncttrue
\mciteSetBstMidEndSepPunct{\mcitedefaultmidpunct}
{\mcitedefaultendpunct}{\mcitedefaultseppunct}\relax
\EndOfBibitem
\bibitem[Matsuo \latin{et~al.}(2023)Matsuo, Imoto, Yokoyama, Sato, Lindemann,
  Gronin, Gardner, Nakosai, Tanaka, Manfra, and Tarucha]{Matsuo2023}
Matsuo,~S.; Imoto,~T.; Yokoyama,~T.; Sato,~Y.; Lindemann,~T.; Gronin,~S.;
  Gardner,~G.~C.; Nakosai,~S.; Tanaka,~Y.; Manfra,~M.~J.; Tarucha,~S.
  Phase-dependent {A}ndreev molecules and superconducting gap closing in
  coherently coupled {J}osephson junctions. arXiv:2303.10540, 2023\relax
\mciteBstWouldAddEndPuncttrue
\mciteSetBstMidEndSepPunct{\mcitedefaultmidpunct}
{\mcitedefaultendpunct}{\mcitedefaultseppunct}\relax
\EndOfBibitem
\bibitem[Cheah \latin{et~al.}(2023)Cheah, Haxell, Schott, Zeng, Paysen, Kate,
  Coraiola, Landstetter, Zadeh, Trampert, Sousa, Riel, Nichele, Wegscheider,
  and Krizek]{Cheah2023}
Cheah,~E.; Haxell,~D.~Z.; Schott,~R.; Zeng,~P.; Paysen,~E.; Kate,~S. C.~t.;
  Coraiola,~M.; Landstetter,~M.; Zadeh,~A.~B.; Trampert,~A.; Sousa,~M.;
  Riel,~H.; Nichele,~F.; Wegscheider,~W.; Krizek,~F. Control over epitaxy and
  the role of the {InAs/Al} interface in hybrid two-dimensional electron gas
  systems. arXiv:2301.06795, 2023\relax
\mciteBstWouldAddEndPuncttrue
\mciteSetBstMidEndSepPunct{\mcitedefaultmidpunct}
{\mcitedefaultendpunct}{\mcitedefaultseppunct}\relax
\EndOfBibitem
\bibitem[Haxell \latin{et~al.}(2023)Haxell, Cheah, K\ifmmode \check{r}\else
  \v{r}\fi{}\'{\i}\ifmmode~\check{z}\else \v{z}\fi{}ek, Schott, Ritter,
  Hinderling, Belzig, Bruder, Wegscheider, Riel, and Nichele]{Haxell2023}
Haxell,~D.~Z.; Cheah,~E.; K\ifmmode \check{r}\else
  \v{r}\fi{}\'{\i}\ifmmode~\check{z}\else \v{z}\fi{}ek,~F.; Schott,~R.;
  Ritter,~M.~F.; Hinderling,~M.; Belzig,~W.; Bruder,~C.; Wegscheider,~W.;
  Riel,~H.; Nichele,~F. Measurements of phase dynamics in planar Josephson
  junctions and SQUIDs. \emph{Phys. Rev. Lett.} \textbf{2023}, \emph{130},
  087002\relax
\mciteBstWouldAddEndPuncttrue
\mciteSetBstMidEndSepPunct{\mcitedefaultmidpunct}
{\mcitedefaultendpunct}{\mcitedefaultseppunct}\relax
\EndOfBibitem
\bibitem[Sup()]{Supplement}
See the Supplemental Material for a detailed discussion on sample fabrication
  and electrical measurements, and for additional experimental results.\relax
\mciteBstWouldAddEndPunctfalse
\mciteSetBstMidEndSepPunct{\mcitedefaultmidpunct}
{}{\mcitedefaultseppunct}\relax
\EndOfBibitem
\bibitem[van Woerkom \latin{et~al.}(2017)van Woerkom, Proutski, van Heck,
  Bouman, V\"{a}yrynen, Glazman, Krogstrup, Nyg{\aa}rd, Kouwenhoven, and
  Geresdi]{vanWoerkom2017}
van Woerkom,~D.~J.; Proutski,~A.; van Heck,~B.; Bouman,~D.;
  V\"{a}yrynen,~J.~I.; Glazman,~L.~I.; Krogstrup,~P.; Nyg{\aa}rd,~J.;
  Kouwenhoven,~L.~P.; Geresdi,~A. Microwave spectroscopy of spinful {A}ndreev
  bound states in ballistic semiconductor {J}osephson junctions. \emph{Nat.
  Phys.} \textbf{2017}, \emph{13}, 876--881\relax
\mciteBstWouldAddEndPuncttrue
\mciteSetBstMidEndSepPunct{\mcitedefaultmidpunct}
{\mcitedefaultendpunct}{\mcitedefaultseppunct}\relax
\EndOfBibitem
\bibitem[Kjaergaard \latin{et~al.}(2017)Kjaergaard, Suominen, Nowak, Akhmerov,
  Shabani, Palmstr\o{}m, Nichele, and Marcus]{Kjaergaard2017}
Kjaergaard,~M.; Suominen,~H.~J.; Nowak,~M.~P.; Akhmerov,~A.~R.; Shabani,~J.;
  Palmstr\o{}m,~C.~J.; Nichele,~F.; Marcus,~C.~M. Transparent
  semiconductor--superconductor interface and induced gap in an epitaxial
  heterostructure {J}osephson junction. \emph{Phys. Rev. Appl.} \textbf{2017},
  \emph{7}, 034029\relax
\mciteBstWouldAddEndPuncttrue
\mciteSetBstMidEndSepPunct{\mcitedefaultmidpunct}
{\mcitedefaultendpunct}{\mcitedefaultseppunct}\relax
\EndOfBibitem
\bibitem[Matsuo \latin{et~al.}(2023)Matsuo, Imoto, Yokoyama, Sato, Lindemann,
  Gronin, Gardner, Manfra, and Tarucha]{Matsuo2023b}
Matsuo,~S.; Imoto,~T.; Yokoyama,~T.; Sato,~Y.; Lindemann,~T.; Gronin,~S.;
  Gardner,~G.~C.; Manfra,~M.~J.; Tarucha,~S. Engineering of anomalous
  {J}osephson effect in coherently coupled {J}osephson junctions.
  arXiv:2305.06596, 2023\relax
\mciteBstWouldAddEndPuncttrue
\mciteSetBstMidEndSepPunct{\mcitedefaultmidpunct}
{\mcitedefaultendpunct}{\mcitedefaultseppunct}\relax
\EndOfBibitem
\bibitem[Annunziata \latin{et~al.}(2010)Annunziata, Santavicca, Frunzio,
  Catelani, Rooks, Frydman, and Prober]{Annunziata2010}
Annunziata,~A.~J.; Santavicca,~D.~F.; Frunzio,~L.; Catelani,~G.; Rooks,~M.~J.;
  Frydman,~A.; Prober,~D.~E. Tunable superconducting nanoinductors.
  \emph{Nanotechnology} \textbf{2010}, \emph{21}, 445202\relax
\mciteBstWouldAddEndPuncttrue
\mciteSetBstMidEndSepPunct{\mcitedefaultmidpunct}
{\mcitedefaultendpunct}{\mcitedefaultseppunct}\relax
\EndOfBibitem
\end{mcitethebibliography}

\clearpage
\newcounter{myc} 
\renewcommand{\thefigure}{S.\arabic{myc}}


\section{Materials and Methods}
The heterostructure used in this work was grown with molecular beam epitaxy techniques on an InP (001) substrate. The top part of the heterostructure consisted of a step-graded InAlAs buffer, and an $8~\mathrm{nm}$ thick InAs quantum well confined between two $\mathrm{In_{0.75}Ga_{0.25}As}$ barriers. The bottom and top barriers were $6~\mathrm{nm}$ and $13~\mathrm{nm}$ thick, respectively. On top of the III--V stack, two monolayers of GaAs and a $15~\mathrm{nm}$ thick Al layer were deposited \textit{in situ}, without breaking vacuum. Characterization of the 2DEG in a Hall bar geometry revealed a peak mobility of $18\times10^3~\mathrm{cm^{2}V^{-1}s^{-1}}$ at an electron sheet density of ${8\times10^{11}~\mathrm{cm^{-2}}}$. This resulted in an electron mean free path $l_{\mathrm{e}}\gtrsim260~\mathrm{nm}$, indicating that Josephson junctions in our devices were in the ballistic regime. The superconducting coherence length in InAs was calculated as $\xi_{\mathrm{InAs}}=\sqrt{\hbar v_{\mathrm{F}}l_{\mathrm{e}}/(2\mathit{\Delta}^*)}=600~\mathrm{nm}$. Here $\hbar$ is the reduced Plank constant, $v_{\mathrm{F}}$ is the electron Fermi velocity, and {$\mathit{\Delta}^*=180~\mathrm{\mu eV}$} is the induced superconducting gap in InAs, which we consider similar to that of bulk Al. Fabrication of the devices was conducted in an identical manner to that described in Ref.~\onlinecite{Coraiola2023}. 

The Al film had a kinetic inductance of $1.7~\mathrm{pH}$ per unit square, calculated from a superconducting Hall bar in the same material~\onlinecite{Annunziata2010,Coraiola2023}. The geometric and kinetic inductance contributions are calculated for the outer [$\PhiL+\PhiR$ in Fig.~1(a) of the Main Text] and inner [$\PhiR$ in Fig.~1(a) of the Main Text] loops. The values for the outer (inner) loop were $30~\mathrm{pH}$ ($15~\mathrm{pH}$) for the geometric inductance and $170~\mathrm{pH}$ ($90~\mathrm{pH}$) for the kinetic inductance. The portion of the circuit shared by both loops, corresponding to the right branch in Fig.~1(a) of the Main Text, had a kinetic inductance of $50~\mathrm{pH}$ and a geometric inductance of $8~\mathrm{pH}$.

Josephson junctions (JJs) were identical in design for all devices. From the junction geometry, the approximate number of transverse modes sustained by the junction is $N\approx W/(\lambda_{\mathrm{F}}/2)$, where $\lambda_{\mathrm{F}}$ is the Fermi wavelength. From gated Hall bar measurements, the sheet carrier density is expected to vary in a range $4${--}$22\times10^{11}~\mathrm{cm^{-2}}$ for typical values of top-gate voltage. This gives a Fermi wavelength of between $17~\mathrm{nm}$ and $40~\mathrm{nm}$, implying a number of modes between $40$ and $100$. Measurements of the current--phase relation (CPR) in JJ1 of Device~1 [Fig.~3(b) of the Main Text] show a non-sinusoidal CPR, indicating the presence of highly transparent modes. The CPR at $V_{1}=0$ [see Fig.~3(c) of the Main Text] gives an effective junction transmission of $\bar{\tau}=0.80$ over an effective number of highly-transmissive modes $\bar{N}=16$~\cite{Nichele2020}. For $V_{1}>-1.5~\mathrm{V}$, $\bar{\tau}$ was approximately constant.

Electrical measurements were performed in a dilution refrigerator with a mixing chamber base temperature below $10~\mathrm{mK}$. Electrical contacts to each device, except for the two flux lines, were provided by resistive looms with QDevil pi-filters at the mixing chamber level and RC filters at both mixing chamber and sample stage. The bias current $I$ passing through the devices was sourced via a Keysight 33600 Waveform Generator. The two output channels produced two synchronized and opposite voltage sawtooth waveforms with amplitude of about $6~\mathrm{V}$ (depending on the specific device) and repetition rate of $133~\mathrm{Hz}$. The two waveforms were applied via two $163~\mathrm{k\Omega}$ resistors placed in series to device source and drain contact, resulting in a maximum current of approximately $35~\mathrm{\mu A}$. The voltage drop $V$ across the device was measured in a four-terminal configuration via a home-made differential amplifier with a gain of 1000, a further amplification stage of 42 provided by the internal amplifier of a Standford Research SR860 lock-in amplifier, and finally detected by a Keysight DSOX2024A oscilloscope. The oscilloscope measured the time needed for the voltage drop across the device to overcome a threshold, set at $7\%$ of the maximum voltage measured in the resistive state. The switching time was averaged over 16 current ramps and converted into a current. With these measurement parameters, transition from superconducting to resistive state was extremely sharp, making the exact choice of the experimental parameters irrelevant.
Flux lines were connected via a superconducting loom, with pi-filters at the mixing chamber level to suppress high-frequency noise, resulting in a total line resistances below $5~\mathrm{\Omega}$. Currents $\IL$ and $\IR$ were generated by two Yokogawa GS200 sources set to current mode. Low-pass RC filters with $R=10~\mathrm{k\Omega}$ and $C=1~\mathrm{\mu F}$ were placed at the current source output.

\section{Current-to-Phase Conversion}
\begin{figure*}
	\includegraphics[width=\textwidth]{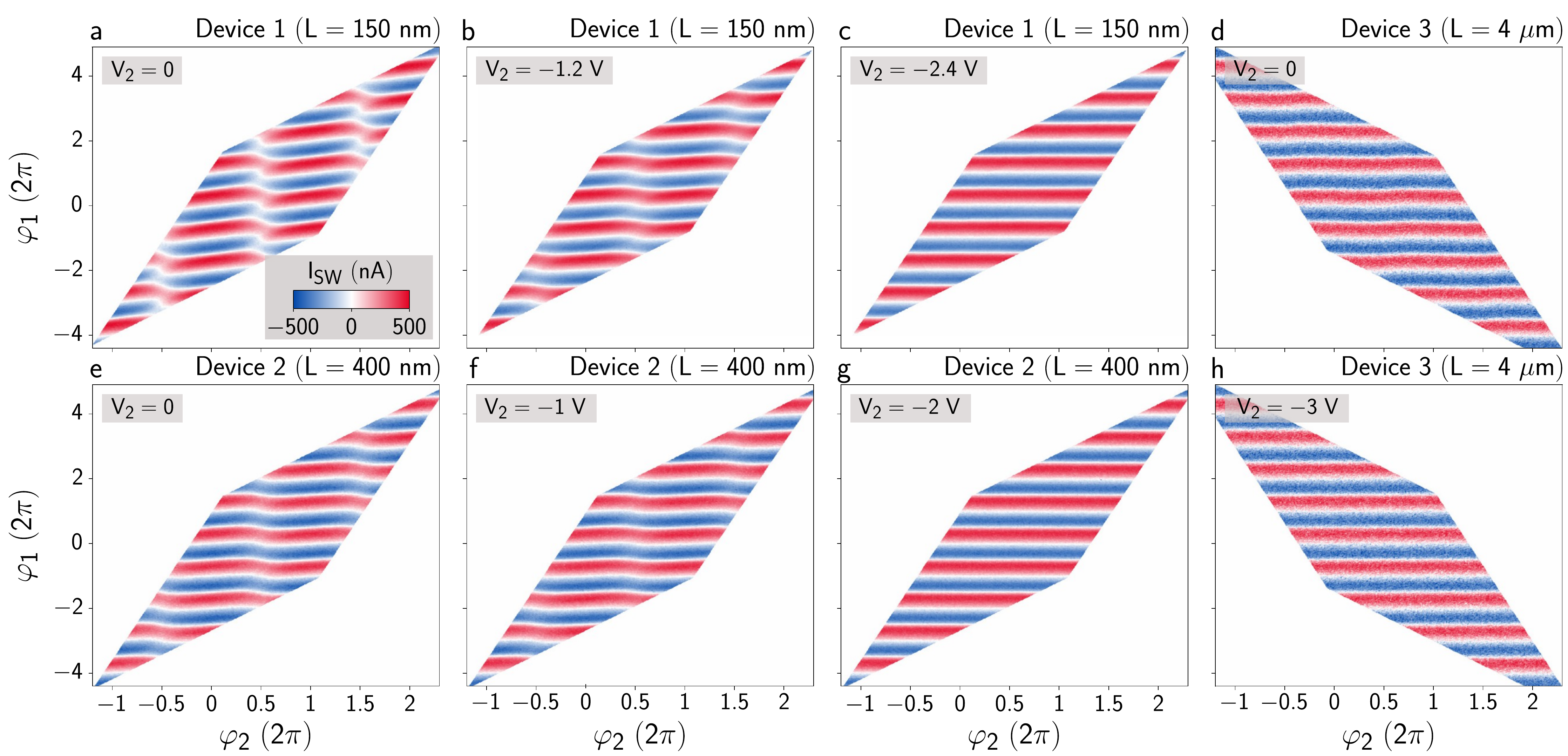}
	\caption{Current-to-phase remapping of switching current measurements. (a-c) Switching current $\Isw$ for Device~1, after subtracting the switching current of the Al constriction $\Ial$, for $V_1=0$ and $V_2=0$, $-1.2$ and $-2.4~\mathrm{V}$, respectively. Data as in Fig.~2(a-c) of the Main Text, plotted as a function of the phase differences across JJ1 and JJ2, respectively $\varphi_1$ and $\varphi_2$. Flux line currents ($\IL,\IR$) were converted to phases ($\varphi_1,\varphi_2$) using Eq.~\ref{eq1}, obtained from the periodicity directions ($\PhiL,\PhiR-\PhiL$). (d) Switching current $\Isw$ of Device~3 for ${V_1=V_2=0}$, in the (${\varphi_1,\varphi_2}$) basis. (e-g) Same as (a-c) for Device~2, with $V_2=0$, $-1$ and $-2~\mathrm{V}$ respectively. (h) Switching current $\Isw$ in Device~3 for ${V_2=-3~\mathrm{V}}$, in the (${\varphi_1,\varphi_2}$) basis. All measurements shown in this figure were taken at $V_1=0$.}
	\label{figS2}
\end{figure*}

As described in the Main Text, currents $\IL$ and $\IR$ were injected into flux bias lines proximal to the device. Each current generated a magnetic field, predominantly impinging on the closest loop: $\IL$ mainly controlled an external flux $\PhiL$ threading the left loop, and $\IR$ mainly controlled an external flux $\PhiR$ through the right loop. Nevertheless, each flux line had a finite coupling to the furthest loop, meaning that $\PhiL$ and $\PhiR$ depended on both $\IL$ and $\IR$. The phase difference across JJ1, $\varphi_1$, changed most strongly as a result of a flux threading the outer loop of the device [see Fig.~1(a) of the Main Text], which corresponds to ${\PhiL+\PhiR}$. This is because a path along the outer loop contains only JJ1 (and the Al constriction), so the phase difference across JJ1 was proportional to the flux ${\PhiL+\PhiR}$ threading that area. We note that the arrow labeled ${\PhiL+\PhiR}$ in Fig.~2(a) of the Main Text corresponds to ${\PhiL+\PhiR}=\Phio$. The phase difference $\varphi_1$ was constant along the ${\PhiR-\PhiL}$ direction, since any increase in the flux through one loop was compensated by the flux through the other. We apply the same procedure to JJ2, showing that the phase $\varphi_2$ across JJ2 varied most strongly as a function of $\PhiR$ and was constant along the $\PhiL$ direction. We define the phase axes as those along which only one phase varies, meaning $\PhiL$ corresponds to $\varphi_1$ and ${\PhiR-\PhiL}$ corresponds to $\varphi_2$. We therefore define these as our phase axes, and perform the conversion from (${\IL,\IR}$) to (${\varphi_1,\varphi_2}$) using the relation:

\begin{equation}\label{eq1}
	\begin{pmatrix}
		\varphi_1 \\
		\varphi_2
	\end{pmatrix}
	\equiv \frac{1}{\Phio}
	\begin{pmatrix}
		\PhiL \\
		\PhiR - \PhiL
	\end{pmatrix} 
	= \frac{1}{\Phio}\mathbf{M}\cdot
	\begin{pmatrix}
		\IL \\
		\IR	
	\end{pmatrix}
	= \frac{1}{\Phio}
	\begin{pmatrix}
		M_{11} & M_{12} \\
		M_{21} & M_{22}
	\end{pmatrix}
	\cdot
	\begin{pmatrix}
		\IL \\
		\IR
	\end{pmatrix},
\end{equation} 
where $\Phio=h/(2e)$ is the superconducting flux quantum and $\mathbf{M}$ is a matrix relating the flux line currents ($\IL,\IR$) to the fluxes (${\PhiL,\PhiR-\PhiL}$). We calculate $\mathbf{M}$ for each device, using the switching current measurements taken at ${V_1=V_2=0}$ [see Figs.~2(a) and (e) of the Main Text for Devices~1 and 2, respectively]. We evaluate Eq.~\ref{eq1} for ${(\PhiL,\PhiR-\PhiL)=(\Phio,0)}$ and ${(\PhiL,\PhiR-\PhiL)=(0,\Phio)}$, and thereby obtain:

\begin{equation}
	\mathbf{M} = 
	\begin{pmatrix}
		0.66 & -1.99\\
		6.14 & -3.00
	\end{pmatrix}~\mathrm{pH}
	\label{eq2}
\end{equation}
for Device~1 and 
\begin{equation}
	\mathbf{M} = 
	\begin{pmatrix}
		0.75 & -2.05 \\
		6.28 & -3.10
	\end{pmatrix}~\mathrm{pH}
	\label{eq2b}
\end{equation}
for Device~2. The good agreement between Eqs.~\ref{eq2} and \ref{eq2b} show that the loop sizes and the flux line fabrication was almost identical between Devices~1 and 2. The matrix $\mathbf{M}$ from Eq.~\ref{eq2} was used for Device~3, where independent evaluation was not possible due to the presence of only two periodicity axes. The position of $\PhiL=\PhiR=0$ was defined where the $\Isw=0$ line intersected all flux periodicity axes, for increasing $\Isw$ in the $\PhiL+\PhiR$ direction. Using the matrix $\mathbf{M}$ of Eq.~\ref{eq2} or~\ref{eq2b}, we apply the linear transformation of Eq.~\ref{eq1} to convert the ${(\IL,\IR)}$ axes to ${(\varphi_1,\varphi_2)}$. The result is plotted in Fig.~\ref{figS2}, for the data presented in Fig.~2 of the Main Text.

\setcounter{myc}{1}

Linecuts of the data in Fig.~\ref{figS2} along $\varphi_1$ are plotted in Fig.~4(a,~d) of the Main Text, at $\varphi_2=0.8\pi$. The anomalous switching current $\Delta I_1$, plotted in Fig.~4(b,~e) of the Main Text, was obtained as a linecut of the data in Fig.~\ref{figS2} along $\varphi_2$, for $\varphi_1=0$. To account for small misalignment of the $\PhiL=\PhiR=0$ origin with respect to the data, a constant switching current offset was subtracted from each dataset such that the oscillations in $\Delta I_1$ were symmetric. Anomalous phase shifts $\Delta\varphi_1$ were calculated from the $\Isw=0$ position where $\partial\Isw/\partial\varphi_1>0$, as a function of $\varphi_2$. Phase shifts $\Delta\varphi_1$ were calculated relative to the data where no current flowed through JJ2 ($V_2=-2.4~\mathrm{V}$ for Device~1, $V_2=-2~\mathrm{V}$ for Device~2). We expect a symmetric deviation in phase across a full period, so a small constant offset was independently obtained and subtracted from each dataset such that the oscillations in $\Delta\varphi_1$ were symmetric.

\section{Measurements on Device~4}
\begin{figure*}
	\includegraphics[width=\textwidth]{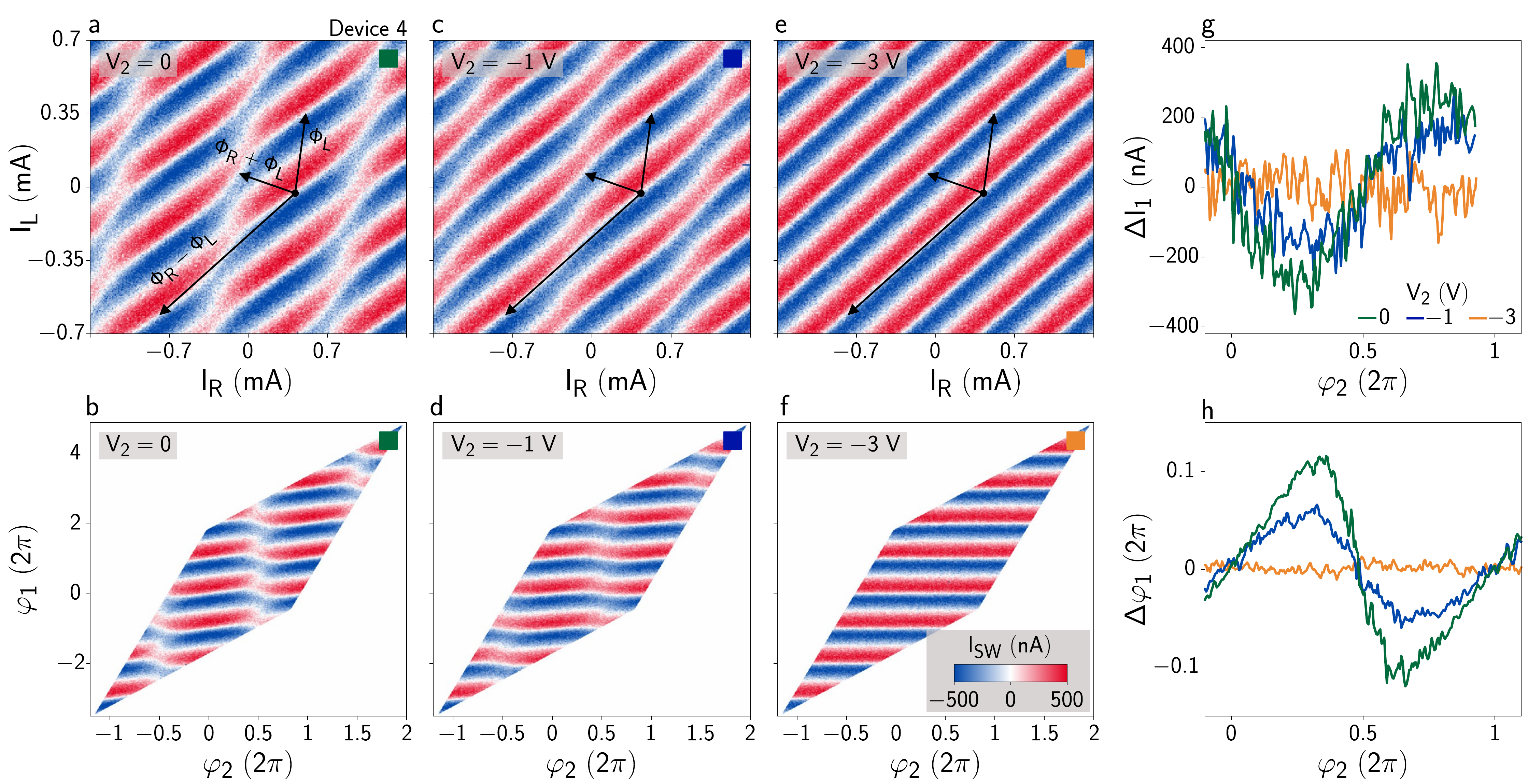}
	\caption{Phase-dependent supercurrent in Device~4. (a,~b) Switching current $\Isw$ of Device~4, identical to Device~1 with a length of $L=150~\mathrm{nm}$, as a function of flux-line currents $(\IL,\IR)$ and junction phases $(\varphi_1,\varphi_2)$, respectively. Data is plotted after subtraction of the switching current of the Al constriction $\Ial$, for $V_1=V_2=0$. (c,~d) Same as (a,~b) for $V_2 = -1~\mathrm{V}$. (e,~f) Same as (a,~b) for $V_2=-3~\mathrm{V}$. (g) Anomalous supercurrent $\Delta I_1$ at $\varphi_1=0$, as a function of $\varphi_2$ for three values of $V_2$ [colors, corresponding to (b,~d,~f)]. (h) Anomalous phase offset $\Delta\varphi_1$ as a function of $\varphi_2$ for three values of $V_2$ [see legend in (g)].}
	\label{figS3}
\end{figure*}

Measurements were performed on a fourth device, identical in design to Device~1 ($L=150~\mathrm{nm}$). Switching current measurements of Device~4 are summarized in Fig.~\ref{figS3}, after subtracting a background corresponding to the switching current of the Al constriction, $\Ial$. The switching current $\Isw$ was measured as a function of the current injected into the left and right flux lines, $\IL$ and $\IR$. The currents ($\IL,\IR$) correspond to fluxes ($\PhiL,\PhiR$) threading the left and right loops, respectively. The phase difference across JJ1, $\varphi_1$, is expected to be modulated most strongly for fluxes threading both loops, i.e., ${\PhiL+\PhiR}$. In the case of no coupling between the JJs, $\varphi_1$ is expected to be constant as a function of $\PhiR-\PhiL$. The phase difference across JJ2 is constant as a function of $\PhiL$. These directions are indicated on Fig.~\ref{figS3}(a) as the black arrows. When $V_1=V_2=0$ [Fig.~\ref{figS3}(a)], there is a clear distortion of the switching current away from the phase axes, indicating hybridization with JJ2. From Fig.~\ref{figS3}(a), the size and shape of this distortion is qualitatively similar to that of Device~1 in the same gate configuration [Fig.~2(a) of the Main Text]. Figure~\ref{figS3}(b) shows the switching current as a function of phase differences across the JJs, (${\varphi_1,\varphi_2}$), obtained using the same method outlined in Eq.~\ref{eq1}. The transformation matrix $\mathbf{M}$ for Device~4 was:

\begin{equation}\label{eq3}
	\mathbf{M} = 
	\begin{pmatrix}
		0.64 & -1.98 \\
		6.23 & -3.00
	\end{pmatrix}~\mathrm{pH},
\end{equation}
very similar to Eqs.~\ref{eq2} and \ref{eq2b} for Devices~1 and 2. 

Measurements were performed for different gate voltages $V_2$ applied to JJ2: $V_2=-1~\mathrm{V}$ for Fig.~\ref{figS3}(c) and $V_2=-3~\mathrm{V}$ for Fig.~\ref{figS3}(e). Figures~\ref{figS3}(d) and (f) show the switching current after transformation by matrix $\mathbf{M}$, for Figs.~\ref{figS3}(c) and (e) respectively. When JJ2 was partially depleted, but still allowed a current to flow, there was a distortion of $\Isw$ but it was less pronounced than for $V_2=0$. For $V_2=-3~\mathrm{V}$, where no current could flow through the fully closed JJ2, there was no distortion of the switching current from the $\PhiR-\PhiL$ direction and oscillations in $\Isw$ occurred with a single periodicity axis. In this configuration, there was no coupling between JJs and the current-phase relation (CPR) was that of JJ1 alone.

The anomalous switching current $\Delta I_1$ at $\varphi_1=0$ is plotted in Fig.~\ref{figS3}(g) as a function of $\varphi_2$, for different gate voltages $V_2$ [colors]. A large, $\varphi_2$-dependent anomalous switching current was observed for $V_2=0$, which was smaller for $V_2=-1~\mathrm{V}$ and absent for $V_2=-3~\mathrm{V}$. The phase shift $\Delta\varphi_1$ is quantified in Fig.~\ref{figS3}(h), as a function of $\varphi_2$ for different gate voltages $V_2$. The maximum phase shift for $V_2=0$ was $\Delta\varphi_1=\pm0.24\pi$ at $\varphi_2=0.8\pi$, almost identical to the result of Device~1. The size of the phase shift was smaller for more negative $V_2$, and completely suppressed when JJ2 was closed.

\setcounter{myc}{2}

\begin{figure*}
	\includegraphics[width=\textwidth]{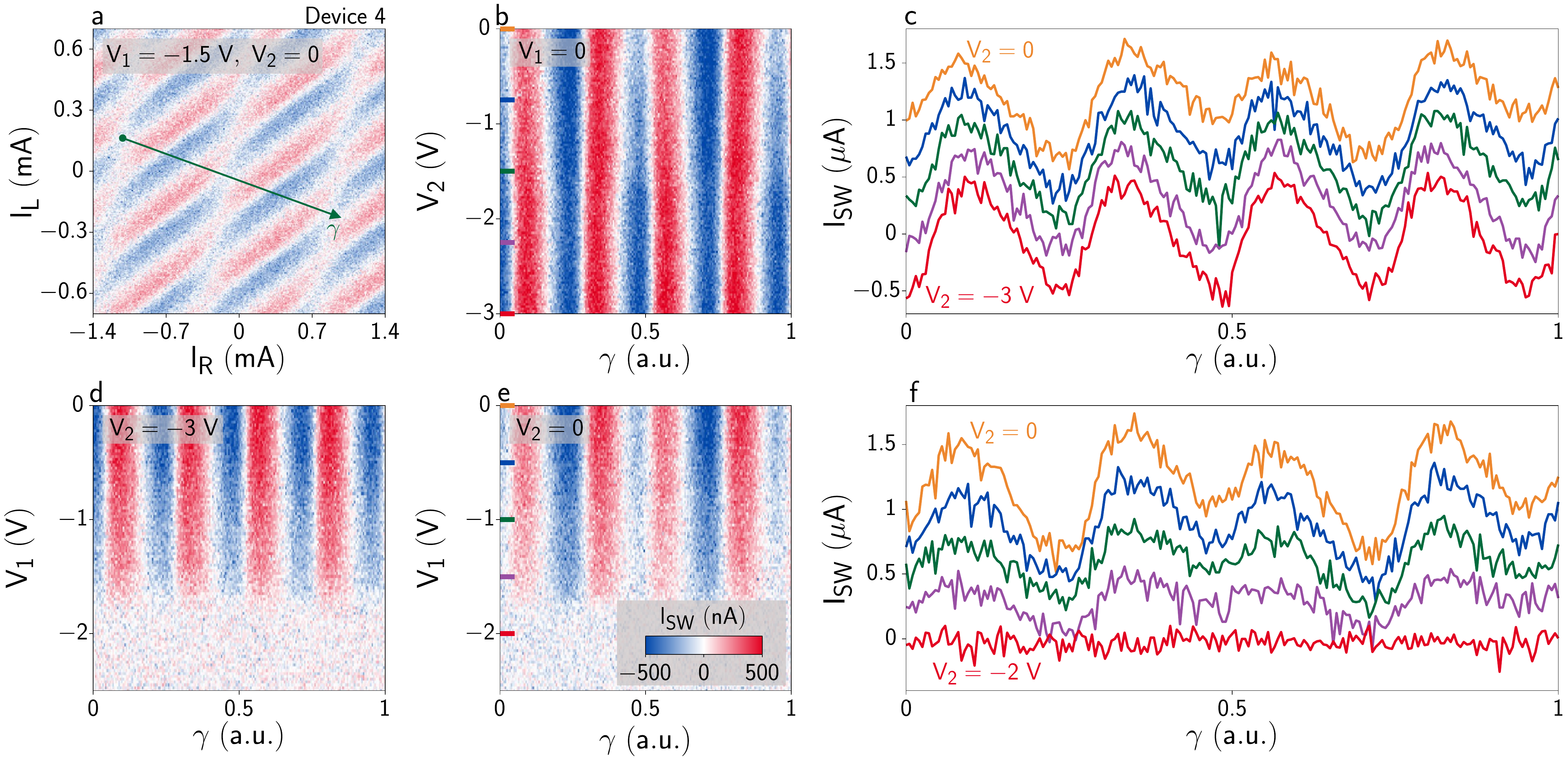}
	\caption{Gate dependence of switching currents in Device~4. (a) Switching current $\Isw$ for Device~4 as a function of flux line currents $\IL$ and $\IR$, at $V_1=-1.5~\mathrm{V}$ and $V_2=0$. Black arrows indicate the direction $\PhiL+\PhiR$, which is the direction of maximum modulation of $\varphi_1$, and $\PhiR-\PhiL$, where $\varphi_1$ is constant and $\varphi_2$ is modulated. Data is qualitatively similar to that of Device~1, in Fig.~3(a) of the Main Text. The path $\gamma$, parallel to the $\PhiL+\PhiR$ direction, is indicated by the arrow. (b) Switching current $\Isw$ along $\gamma$, as a function of $V_2$ with $V_1=0$. In this configuration, the positions where $\Isw=0$ shift as $V_2$ is decreased. (c) Linecuts of $\Isw$ extracted from (b), at different values of $V_2$ [indicated by the colored markers in (b)]. (d) Switching current $\Isw$ along $\gamma$ as a function of $V_1$, with $V_2=-3~\mathrm{V}$ such that no current flows through JJ2. (e) As in (d), but measured with $V_2=0$. Also in this configuration, despite a current can flow through both JJs, there is no shift in the $\Isw=0$ position. (f) Linecuts of $\Isw$ extracted from (e), at different values of $V_1$ [indicated by the colored markers in (e)].}
	\label{figS1}
\end{figure*}

Figure~\ref{figS1} presents the dependence of the switching current of Device~4 on the gate voltages $V_1$ and $V_2$. Figure~\ref{figS1}(a) shows $\Isw$ as a function of flux line currents $\IL$ and $\IR$, for $V_1=-1.5~\mathrm{V}$ and $V_2=0$. While the oscillation amplitude was reduced relative to the $V_1=0$ configuration, the switching current modulation was comparable to that of Fig.~\ref{figS3}(a). The switching current was measured along the path $\gamma$, parallel to the $\PhiL+\PhiR$ direction. Figure~\ref{figS1}(b) shows $\Isw$ along $\gamma$ as a function of $V_2$, with $V_1=0$. Selected linecuts are shown in Fig.~\ref{figS1}(c). The position of $\Isw=0$ shifted as a function of $V_2$ and the oscillations changed from being distorted for $V_2>-1.5~\mathrm{V}$, to a regularly skewed CPR for $V_2<-1.5~\mathrm{V}$. This demonstrates the strong effect of $V_2$ on the CPR of JJ1. Figures~\ref{figS1}(d) and (e) show $\Isw$ along $\gamma$ as a function of $V_1$, for $V_2=-3~\mathrm{V}$ and $0$ respectively. The switching current had a conventional forward-skewed CPR in Fig.~\ref{figS1}(d), since no current flowed through JJ2. Decreasing $V_1$ only decreased the amplitude of oscillations. For $V_2=0$, a current flowed through JJ2. Nevertheless, $V_1$ only caused a decrease in the overall amplitude of oscillations, without introducing distortions or phase shifts. This is also evident from selected linecuts, displayed in Fig.~\ref{figS1}(f), showing that $V_1$ had little to no effect on the anomalous phase shift $\Delta\varphi_1$.

\setcounter{myc}{3}


\end{document}